\documentclass[fleqn,usenatbib,useAMS]{mnras}

\usepackage{newtxtext,newtxmath}

\usepackage[T1]{fontenc}
\usepackage{ae,aecompl}

\DeclareRobustCommand{\VAN}[3]{#2}
\let\VANthebibliography\thebibliography
\def\thebibliography{\DeclareRobustCommand{\VAN}[3]{##3}\VANthebibliography}

\usepackage{graphicx}	
\usepackage{amsmath}	
\usepackage{multicol}        
\usepackage{subcaption}
\usepackage{graphicx}
\usepackage{ulem}
\usepackage{booktabs}
\usepackage{float}
\usepackage{xcolor}
\usepackage{cases} 
\usepackage{caption} 

\title[Inner edges of planetesimal belts]{Inner edges of planetesimal belts: collisionally eroded or truncated?}

\author[A. Imaz Blanco et al.]{Amaia Imaz Blanco$^{1,2}$\thanks{E-mail: aimazblanco@gmail.com},
Sebastian Marino$^{2,3,4}$\thanks{E-mail:sebastian.marino.estay@gmail.com},
Luca Matr\`a$^{5}$, 
Mark Booth$^{6}$,
John Carpenter$^{7}$,
\newauthor Virginie Faramaz$^{8}$,
Thomas Henning$^{9}$,
A. Meredith Hughes$^{10}$,
Grant M. Kennedy$^{11}$,
Sebasti\'an P\'erez$^{12,13,14}$,
\newauthor Luca Ricci$^{15}$ and
Mark C. Wyatt$^{2}$\\
$^{1}$ Physics Department, Lancaster University, Bailrigg, Lancaster, UK \\
$^{2}$ Institute of Astronomy, University of Cambridge, Madingley Road, Cambridge, UK\\
$^{3}$ Jesus College, University of Cambridge, Jesus Lane, Cambridge CB5 8BL, UK\\
$^{4}$School of Physics and Astronomy, University of Exeter, Stocker Road, Exeter, EX4 4QL, UK\\
$^{5}$ School of Physics, Trinity College Dublin, the University of Dublin, College Green, Dublin 2, Ireland \\
$^{6}$ Astrophysikalisches Institut and Universitatssternwarte, Friedrich-Schillar-Universitat, Schillergasschen 2-3, D-7745 Jena, Germany \\
$^{7}$ Joint ALMA Observatory, Alonso de Córdova 3107, Vitacura, Santiago, 763 0355, Chile\\
$^{8}$ Steward Observatory, Department of Astronomy, University of Arizona, 933 N. Cherry Ave, Tucson, AZ 85721, USA \\
$^{9}$ Max Planck Institute for Astronomy, Konigstuhl 17, 69117 Heidelberg, Germany \\
$^{10}$ Wesleyan University, Van Vleck Observatory, 96 Foss Hill Dr, Middletown, CT 06459, USA \\
$^{11}$ Department of Physics, University of Warwick, Coventry CV4 7AL, UK \\
$^{12}$ Departamento de Física, Universidad de Santiago de Chile, Av. Victor Jara 3659, Santiago \\ 
$^{13}$ Millennium Nucleus on Young Exoplanets and their Moons (YEMS), Chile\\
$^{14}$ Center for Interdisciplinary Research in Astrophysics and Space Exploration (CIRAS), Universidad de Santiago de Chile, Estación Central, Chile \\
$^{15}$Department of Physics and Astronomy, California State University Northridge, 18111 Nordhoff Street, Northridge, CA 91330, USA
}

\date{}

\pubyear{2022}

\begin{document}
\label{firstpage}
\pagerange{\pageref{firstpage}--\pageref{lastpage}}
\maketitle

\begin{abstract}
The radial structure of debris discs can encode important information about their dynamical and collisional history. In this paper we present a 3-phase analytical model to analyse the collisional evolution of solids in debris discs, focusing on their joint radial and temporal dependence. Consistent with previous models, we find that as the largest planetesimals reach collisional equilibrium in the inner regions, the surface density of dust and solids becomes proportional to $\sim r^{2}$ within a certain critical radius. We present simple equations to estimate the critical radius and surface density of dust as a function of the maximum planetesimal size and initial surface density in solids (and vice versa). We apply this model to ALMA observations of 7 wide debris discs. We use both parametric and non-parametric modelling to test if their inner edges are shallow and consistent with collisional evolution. We find that 4 out of 7 have inner edges consistent with collisional evolution. Three of these would require small maximum planetesimal sizes below 10~km, with HR~8799's disc potentially lacking solids larger than a few centimeters. The remaining systems have inner edges that are much sharper, which requires maximum planetesimal sizes $\gtrsim10$~km. Their sharp inner edges suggest they could have been truncated by planets, which JWST could detect. In the context of our model, we find that the 7 discs require surface densities below a Minimum Mass Solar Nebula, avoiding the so-called disc mass problem. Finally, during the modelling of HD~107146 we discover that its wide gap is split into two narrower ones, which could be due to two low-mass planets formed within the disc.

\end{abstract}

\begin{keywords}
submillimetre: planetary systems -- planetary systems -- circumstellar matter
\end{keywords}



\section{Introduction} \label{sec:intro}

Debris discs, extrasolar analogues of the asteroid and Kuiper belt, are a ubiquitous component of planetary systems \citep{Wyatt2008, Hughes2018, Marino2022}. These discs are made of solids whose sizes span ten orders of magnitude - from km-sized or larger down to micron-sized. These grind down into a collisional cascade, producing dust that is readily detected as infrared excesses around 20-30\% of nearby AFGK-type stars \citep{Su2006, Sibthorpe2018}. Dozens of discs have been imaged in the optical and NIR tracing $\mu$m-sized grains scattering stellar light \citep[e.g.,][]{Mouillet1997, Milli2017, Feldt2017, Esposito2020} and at millimetre wavelengths tracing the thermal emission of larger mm-sized grains \citep[e.g.,][]{Macgregor2013, Marino2016}. The latter are unperturbed by radiation and gas-drag forces, and thus mm-sized grains tend to trace better the distribution of planetesimals \citep{Thebault2012}.

\begin{figure*}
\vspace{-0mm}
 \centering
   \includegraphics[width=1.0\textwidth]{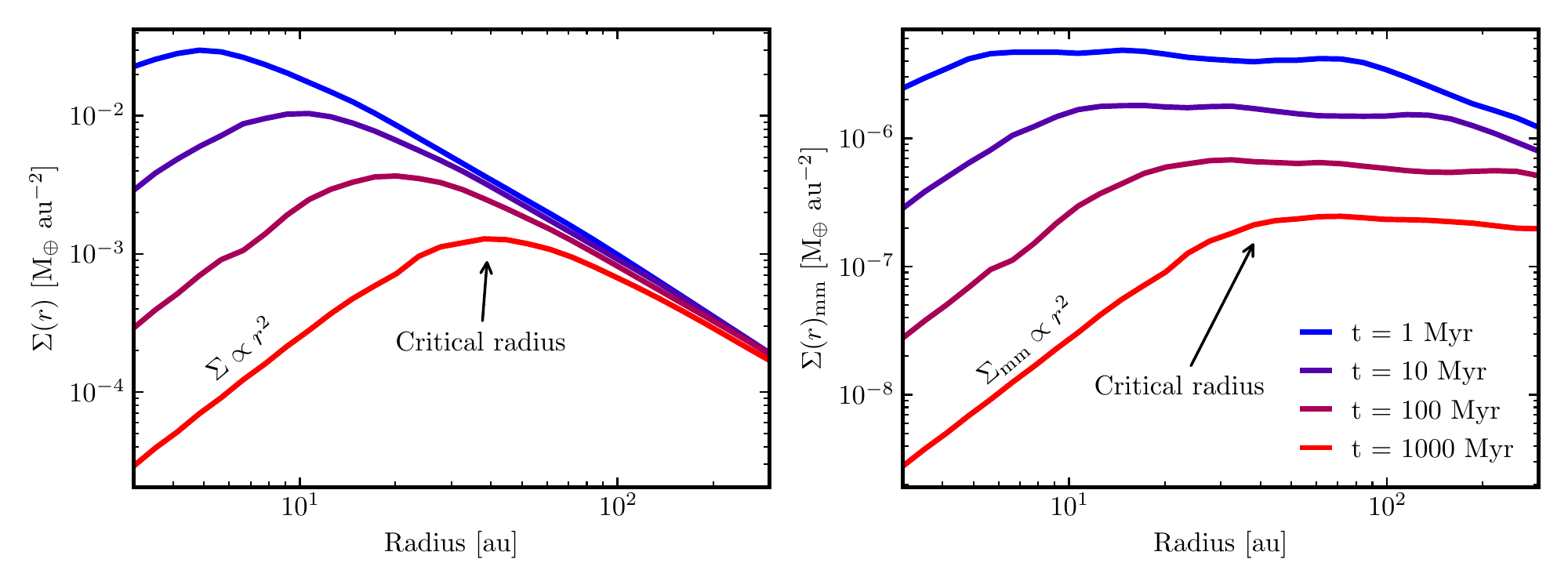}
\caption{Collisional evolution of the surface density of solids (left) and dust smaller than 1~cm (right) as a function of radius. The curves are computed using the numerical model in \citet{Marino2017b} for a 1~$M_{\odot}$ star surrounded by a debris disc with an initial surface density of $(r/1\ \mathrm{au})^{-3/2}\ M_{\oplus}$~au$^{-2}$, a maximum planetesimals diameter of 100~km and the solid strength for ice as in \S\ref{sec:analytical}. The arrows point at the critical radius, which shifts in time towards large radii. The region interior to the critical radius has a surface density of solids and dust roughly proportional to $r^2$.} 
\label{fig:surface_density_model}
\end{figure*}

Imaging debris discs has proven to be a powerful tool for constraining the dynamics and architectures of planetary systems. Disc images reveal their morphology, which can be linked with the presence or absence of shepherding planets. For example, the presence of a warp in $\beta$~Pic's disc hinted at the presence of a massive planet that was later discovered \citep{Mouillet1997, Lagrange2009}. Other discs show evidence of gaps that could have been cleared by planets \citep[e.g.][]{Marino2020hd206, Nederlander2021}, eccentric rings possibly forced by eccentric planets \citep{Kalas2005, Faramaz2019}, clumps that could be due to resonant trapping \citep{Wyatt2006, Dent2014, Han2023}, complex vertical structures that hint at multiple dynamical populations of planetesimals \citep{Matra2019b}, and shallow outer edges that suggest high degrees of dynamical excitation \citep{Marino2021}. Most of the time, these morphologies could be produced by planets smaller than a few Jupiter masses at tens of au that pre-JWST instrumentation was unable to detect \citep{Pearce2022}; $\beta$~Pic~b is an exception in that regard. Therefore, in addition to constraining planetary systems' dynamics, the discs' morphologies serve as an indirect way to infer the presence of planets.    

One feature that is of particular interest is the inner edge of a debris disc. If the outer Solar System architecture was the norm, we would expect debris disc inner edges to be truncated by planets. The inner edge location has been extensively used to infer the location and mass of such planets \citep[e.g.][]{Quillen2006, Chiang2009, NesvoldKuchner2015a, Pearce2022}. However, only for a few systems has the inner edge shape been directly compared to dynamical models to truly assess this scenario \citep[e.g.][]{Chiang2009, Read2018}. Such comparisons require high-resolution and sensitivity observations that have only become available in the last few years with the Atacama Large Millimeter/submillimeter Array (ALMA).

The observed inner edge of discs could, on the other hand, be a result of collisional evolution. As the collisional lifetime of solids decreases with decreasing radius, the inner regions of a wide disc will collisionally deplete faster and become fainter than at larger radii \citep{Kenyon2002, Krivov2006, Wyatt2007a, Kobayashi2010}. This will lead to a surface density that rises with radius up to a critical radius, at which the collisional lifetime of the largest planetesimals and the age of the system (or the time since it was stirred) are the same. Disc observations that are limited by their low resolution or sensitivity might easily miss the low-level parts of the inner regions, and thus misinterpret this critical radius as the disc inner edge. In this scenario, the inner edge would be shallow with a surface density approximately proportional to $r^{7/3}$ \citep{Kennedy2010}. So far, this behaviour of the surface density has been characterized using very simple analytical models that assume either size independent disruption threshold \citep{Kennedy2010} or very narrow debris rings \citep{Lohne2008, Geiler2017}. More complex numerical models of wide debris discs that account for how the strength of solids varies with size have shown this behaviour \citep{Schuppler2016, Marino2017b}, but they did not provide simple equations that characterize the surface density of dust and that could be applied to observations.


In this paper, we present a 3-phase analytical model to characterise the surface density of mm-sized dust undergoing collisional evolution in a wide debris disc and investigate whether such evolution is consistent with the inner edge sharpness that we measure in seven wide debris discs: HD~107146, HD~92945, HD~206893, q$^{1}$~Eri, 49~Ceti, AU~Mic and HR~8799. These seven discs have been well resolved with ALMA, with observations that resolve their radial extent with $>5$ resolution elements and signal-to-noise ratios larger than 10, and thus ideal for investigating their inner edges. This paper is structured as follows. In \S\ref{sec:analytical} we briefly summarise previous collisional models and present our analytical model and relevant equations that can be applied to observations. In \S\ref{sec:dataconstraints} we fit the ALMA data of seven discs to determine if they have sharp inner edges or are rather consistent with having a smooth rising surface density due to collisional evolution. In \S\ref{sec:discussion} we use our collisional model to interpret the results from fitting the ALMA data and discuss our results. Finally, in \S\ref{sec:concl} we summarise our conclusions.

\section{A collisionally eroded disc inner edge} \label{sec:analytical}

The collisional evolution of debris discs is a topic that has been studied at great length to interpret observations of debris discs. Collisional models tend to split into a few different kinds. First, there are simple analytical models that assume a pre-stirred disc with a wide size distribution up to planetesimal sizes and described by a single power law with an exponent of -3.5 \citep[e.g.][]{Dohnanyi1969, DominikDecin2003, Wyatt2007a}. These models were later updated to consider that A) the internal strength of a solid, affecting its collisional lifetime, is a function of the solid's size \citep{OBrien2003}, and B) not all solids have collided by the age of the system, with solids above a certain size retaining a primordial size distribution \citep[e.g.][]{Lohne2008, Shannon2011, Geiler2017}. Including these effects in these analytical models modifies the single power law to a set of up to three power laws, which we will explore in this paper.

A second type of model has still assumed a pre-stirred disc but they numerically solve the size distribution evolution due to collisions and considering additional effects such as radiation pressure and PR-drag \citep[e.g.][]{Krivov2006, Thebault2007, Wyatt2011, Gaspar2012}. These models tend to have a fixed maximum planetesimal size as there is no growth. \citet{Marino2017b} explored this kind of model, for example, to explain how collisional evolution alone may explain the 61 Vir disc's observed flat surface density distribution and inner edge location.

A third type of model has modelled debris discs being born with solids up to ${\sim}1$~km in size and with very small eccentricities and inclinations (dynamically cold). In these models, solids grow through collisions at low relative velocities until the formation of Pluto-sized objects that effectively stir the disc triggering a collisional cascade \citep[e.g.][]{Kenyon2004, Kenyon2008, Kenyon2010, Kobayashi2010, Kobayashi2014}. More recent updates to these models have considered alternative initial conditions where debris discs are born as a mix of cm-sized pebbles and 100~km-sized planetesimals, closer to what could be expected if planetesimals are formed via the streaming instability \citep{Najita2022}. This highlights the uncertain initial conditions of debris discs as we we do not know how debris discs transition from protoplanetary discs \citep{Wyatt2015}. Finally, a fourth type of model has combined N-body simulations and collisional evolution to study the dust production in planet-disc interaction scenarios \citep[e.g.][]{Jackson2012, Kral2013, NesvoldKuchner2015a}.

All these models have shown in one way or another how as a debris disc collisionally evolves, its inner regions will deplete faster. In Figure \ref{fig:surface_density_model} we show this effect using the second type of model as implemented in \cite{Marino2017b} assuming a pre-stirred disc with planetesimals up to 100~km sizes. As the disc evolves, the surface density in the inner regions becomes a simple power law roughly proportional to $r^2$ up to a critical radius $r_{\rm c}$.
This critical radius may be interpreted as the disc's inner edge if observations are unable to resolve and detect the lower densities at smaller radii. One way to assess whether the observed inner edge corresponds to this critical radius is to measure the slope of the surface brightness or density just interior to the "observed" inner edge and compare it with collisional models. If consistent, then the location of the critical radius or observed inner edge together with the dust density at that distance can be used to constrain the maximum planetesimal size feeding the collisional cascade and the initial surface density of solid material \citep{Marino2017b}.

In the following sections, we will present an analytical model inspired by \citet{Lohne2008}, to describe the collisional evolution of an axisymmetric, radially wide, and vertically thin debris disc (with a vertical aspect ratio $h\ll1$) as a function of radius ($r$ in cylindrical coordinates). The key difference compared to previous analytical models is that we will focus on the joint radial and temporal dependence of the disc evolution. In particular, we will:
\begin{itemize}
    \item define an analytical, general 3-phase collisional cascade, assuming no solid growth and motivated by the behaviour of the solids' disruption threshold strength as a function of size (\S\ref{sec:sizedist});
    \item derive an expression for the largest bodies participating in the cascade and their collision timescale, showing how their dependence on radius naturally gives rise to the critical radius and a multi-phase radial distribution of solids (\S\ref{sec:collrate});
    \item derive a full expression for the critical radius, and the expected dust mass surface density at this critical radius, as key observables to infer the maximum planetesimal size and the total surface density (mass) of solids in the planetesimal belt (\S\ref{sec:rcequations});
    \item show that interior to this critical radius, the surface density of a collisionally evolving planetesimal belt should always follow a ${\sim}r^2$ dependence rising up to the critical radius (\S\ref{sec:radialslope}).
\end{itemize}

\subsection{An analytical, 3-phase size distribution approach}
\label{sec:sizedist}
\begin{figure*}
\vspace{-0mm}
 \centering
 \hspace{-3mm}
   \includegraphics*[scale=0.44]{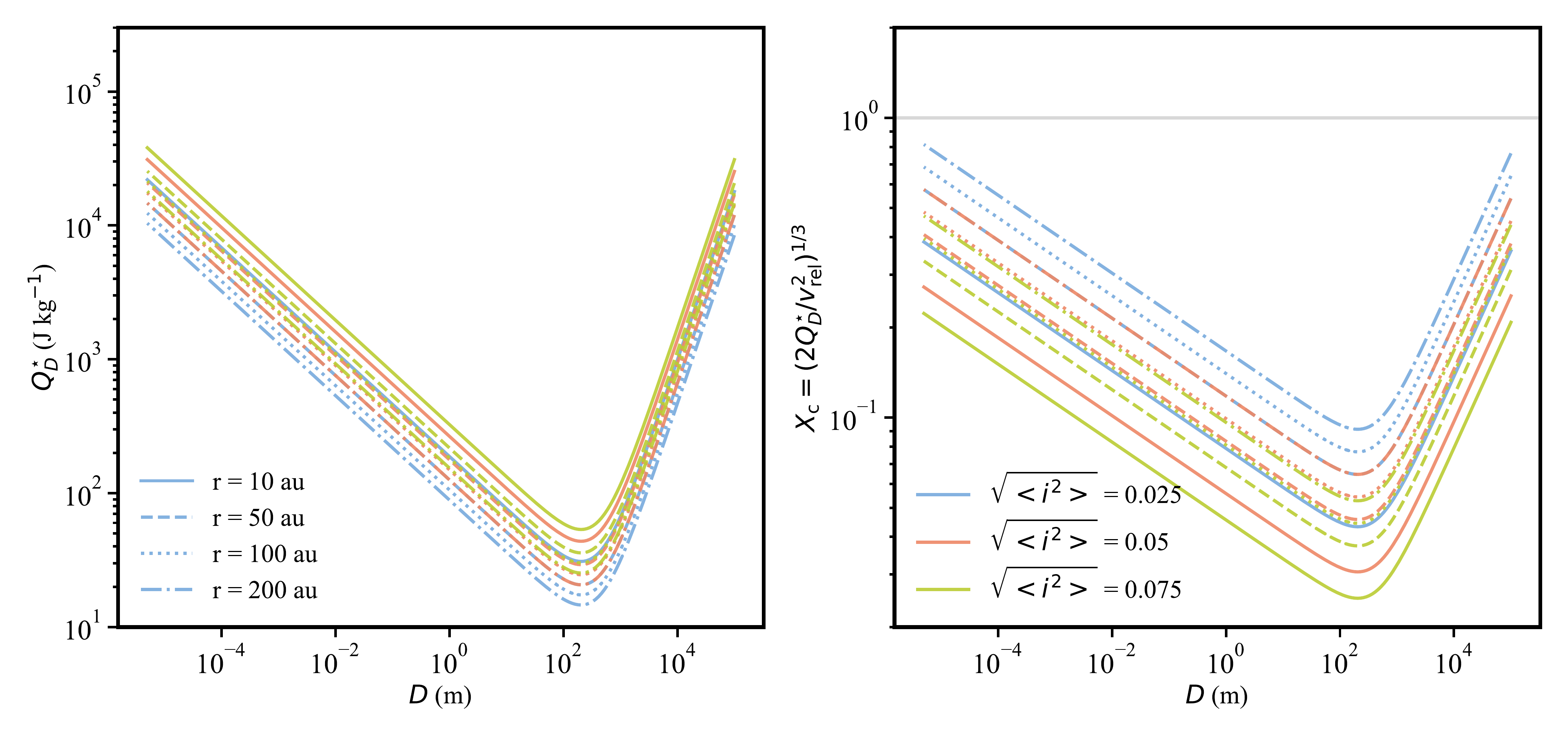}
\vspace{-6mm}
\caption{\textit{Left:} Catastrophic disruption threshold strength $Q_{\rm D}^{\star}$ as a function of size $D$ adopted in our calculations, with different line styles representing different orbital radii and different colors representing different rms inclinations. The curve is well approximated by two power laws shortward (strength regime, with slope $b_{\rm s}$) and longward (gravity regime, with slope $b_{\rm g}$) of size $D_{\rm b}\sim420$ m. \textit{Right:} Minimum impactor size to cause a catastrophic collision, as a fraction ($X_{\rm c}$) of the size of a given target ($D$). Line styles and colors have the same meaning as in the left panels. Targets which have $X_{\rm c}=1$ (grey line) or higher can only be destroyed by impactors of the same size or larger.} 
\label{fig:qdstarxc}
\end{figure*}

We start by approximating the size distribution of solids with an analytical three-phase distribution between minimum grain diameter $D_{\rm min}$ and maximum planetesimal diameter $D_{\rm max}$ as explored e.g. by \citet{Lohne2008}. This takes the form
\begin{equation}
n(D)=n_{D_{\rm max}}\left(\frac{D}{D_{\rm max}}\right)^{2-3q_{\rm p}}\\ \mathrm{for}\ \ D_{\rm c}<D<D_{\rm max}
\label{eq:sizedist1}
\end{equation}
\begin{equation}
n(D)=n_{D_{\rm max}}\left(\frac{D_{\rm c}}{D_{\rm max}}\right)^{2-3q_{\rm p}}\left(\frac{D}{D_{\rm c}}\right)^{2-3q_{\rm g}}\\ \mathrm{for}\ \ D_{\rm b}<D<D_{\rm c}
\label{eq:sizedist2}
\end{equation}
\begin{equation}
n(D)=n_{D_{\rm max}}\left(\frac{D_{\rm c}}{D_{\rm max}}\right)^{2-3q_{\rm p}}\left(\frac{D_{\rm b}}{D_{\rm c}}\right)^{2-3q_{\rm g}}\left(\frac{D}{D_{\rm b}}\right)^{2-3q_{\rm s}}\\ \mathrm{for}\ \ D_{\rm min}<D<D_{\rm b},
\label{eq:sizedist3}
\end{equation}
where $n(D)dD$ is the number of objects with diameters in the range $D$ to $D+dD$.
At the top of the size distribution (Eq. \ref{eq:sizedist1}), objects whose collision timescale ($\tau_{\rm col}$) is longer than the system age ($t_{\rm age}$) have not collided yet, and follow a primordial size distribution assumed to be a power law with slope $2-3q_{\rm p}$ down to objects of size $D_{\rm c}=D(\tau_{\rm col}=t_{\rm age})$, whose collision timescale is equal to the age of the system. Smaller objects are part of the collisional cascade, and the slope of their size distribution (Eq. \ref{eq:sizedist2} and \ref{eq:sizedist3}) arises from catastrophic collisions having disruption threshold strengths $Q^{\star}_{D}$, closely following a double power law with slopes $b_{\rm g}=(11-6q_{\rm g})/(q_{\rm g}-1)$ and $b_{\rm s}=(11-6q_{\rm s})/(q_{\rm s}-1)$ \citep{OBrien2003}, where $q_{\rm g}$ and $q_{\rm s}$ are the resulting size distribution slopes in the strength and gravity regimes, respectively. The boundary between the two slopes takes place at size $D_{\rm b}$ with threshold $Q^{\star}_{D_{\rm b}}$, where we therefore expect a break from $q_{\rm g}$ to $q_{\rm s}$ in the slope of the size distribution. We use the strength law adopted by e.g. \citet{Marino2017b} (their Eq. 2) which has a dependence on the relative velocity of collisions $v_{\rm rel}$, inspired by the results of \citet{BenzAsphaug1999} and \citet{LeinhardtStewart2012}. This takes the form
\begin{equation}
Q^{\star}_{\rm D}=\left[Q_{\rm D,s}\left(\frac{D}{1\rm m}\right)^{b_{\rm s}}+Q_{\rm D,g}\left(\frac{D}{1\rm m}\right)^{b_{\rm g}}\right]\left(\frac{v_{\rm rel}}{v_0}\right)^{\frac{1}{2}},
\label{eq:qdstarlaw}
\end{equation}
which we approximate as
\begin{equation}
Q^{\star}_{\rm D}=Q^{\star}_{D_{\rm b}}\left(\frac{v_{\rm rel}}{v_0}\right)^{\frac{1}{2}}\left(\frac{D}{D_{\rm b}}\right)^{b_{\rm g}}\ \ \ \ \ \ \mathrm{for}\ \ D>D_{\rm b}
\label{eq:qdstarlaw1}
\end{equation}
\begin{equation}
Q^{\star}_{\rm D}=Q^{\star}_{D_{\rm b}}\left(\frac{v_{\rm rel}}{v_0}\right)^{\frac{1}{2}}\left(\frac{D}{D_{\rm b}}\right)^{b_{\rm s}}\ \ \ \ \ \ \mathrm{for}\ \ D<D_{\rm b},
\label{eq:qdstarlaw2}
\end{equation}
where $D_{\rm b}=420$ m, $Q^{\star}_{D_{\rm b}}=33$ J kg$^{-1}$, $b_{\rm s}=-0.39$ (implying $q_{\rm s}=1.89$), $b_{\rm g}=1.26$ (implying $q_{\rm g}=1.69$), and $v_0=3000$ m s$^{-1}$ \citep[consistent with icy solids in simulations by][]{BenzAsphaug1999}. Fig. \ref{fig:qdstarxc} (left) shows the dependence of our adopted $Q_{\rm D}^{\star}$ law on the radius $r$ and inclination rms $i$ of the belt, assuming a 1~$M_{\odot}$ star and an eccentricity rms ($e$) that is twice the inclination rms ($i$). Table~\ref{tab:collpar} summarises the adopted collisional parameters.

\begin{table}
    \centering
    \caption{Collisional parameters that determine the size distribution. These are consistent with ice in simulations by \protect\cite{BenzAsphaug1999}.}
   \setlength\tabcolsep{1.5pt}
    \begin{tabular}{l|l|l}
        \hline
        Parameter & Value & Description \\
        \hline
        $D_{\rm b}$ & 420 m & Boundary size between strength and gravity regimes. \\
        $Q^{\star}_{D_\mathrm{b}}$ & 33 J kg$^{-1}$ & Disruption threshold at size $D_{\rm b}$.\\
        $b_{\rm g}$ & 1.26 & $Q^{\star}_{D}$ slope in the gravity regime. \\  
        $b_{\rm s}$ & -0.39 & $Q^{\star}_{D}$ slope in the strength regime. \\  
        $q_{\rm g}$ & 1.69 & $n(D>D_{\rm b})\propto D^{2-3q_{\rm g}}\propto M^{-q_{\rm g}}$. \\  
        $q_{\rm s}$ & 1.89 & $n(D<D_{\rm b})\propto D^{2-3q_{\rm s}}\propto M^{-q_{\rm s}}$. \\  
        $v_0$       & 3.0 km s$^{-1}$  & Reference relative velocity. \\
        $\rho$      & 1000 kg~m$^{-3}$ & Bulk density of solids. \\
        $i$         & 0.025 & Inclination dispersion (rms). \\
        $e$         & 0.05  & Eccentricity dispersion (rms).
    \end{tabular}
    \label{tab:collpar}
\end{table}

\subsection{The collision rate of the largest bodies in the cascade}
\label{sec:collrate}
We now proceed to derive simple analytical equations for the catastrophic collision rate of solids. For an object of size $D$, collisions with impactors of size $D_{\rm im}$ are only catastrophic if the impactors' specific energy is above $Q^{\star}_\mathrm{D}$. That minimum specific energy can be translated to a minimum size $X_{\rm c}D$ (with $X_{\rm c}\equiv(2Q_{\rm D}^{\star}/(v_{\rm rel}^2))^{1/3}$) since the relative velocities are independent of size here. Therefore, the catastrophic collision rate of material of size $D$ in the absence of gravitational focusing, can be expressed as \citep[e.g.][]{Wyatt2002}
\begin{equation}
R_{\rm col}=\frac{v_{\rm rel}}{V}\int_{X_{\rm c}D}^{D_{\rm max}}n(D_{\rm im})\sigma_{D_{\rm im}}\left(1+\frac{D}{D_{\rm im}}\right)^2dD_{\rm im},
\label{eq:Rcolbasic}
\end{equation}
where $V$ is the volume available for collisions, $n(D_{\rm im})$ is the size distribution, and $\sigma_{D_{\rm im}}=\frac{\pi D_{\rm im}^2}{4}$ is the geometric cross section of a given impactor. Fig. \ref{fig:qdstarxc} (right) shows how the minimum impactor size for a catastrophic collision ($X_{\rm c}$, as a fraction of the target size) varies as a function of size $D$ assuming a $2~M_{\odot}$ star (note that $X_{\rm c}$ is proportional to $1/\sqrt{v_{\rm rel}}$, hence $X_{\rm c}$ has a weak dependence on $M_\star$). For the chosen composition, and all sizes considered here ($D\leq100$~km), we are in the regime where $X_{\rm c}<1$, i.e. the smallest impactors able to destroy a target are smaller than the target itself. This is important since the collisional rate of bodies of size $D$ is typically dominated by the smallest sizes able to disrupt it, i.e. those with a size close to $X_{\rm c}D$.

A fundamental parameter that sets the evolution, size distribution, total mass, and radial distribution of material in a collisional cascade is $D_{\rm c}$, the size whose collisional lifetime is equal to the age of the system (i.e. the timescale at which it experiences a catastrophic collision). Using only Eq. \ref{eq:Rcolbasic}, we can deduce that the collision rate increases with decreasing disc radius.
This is because higher Keplerian velocities and smaller volumes produce more collisions in the inner regions ($v_{\rm rel}\propto v_{k}\propto r^{-0.5}$, and $V=4\pi r^3\frac{dr}{r}I\propto r^3$).
Therefore, the maximum size $D_{\rm c}$ to have suffered at least one collision within the age of the system is larger at smaller radii, and decreases at larger radii. But if $D_{\rm c}$ changes with radius, the size distribution changes with radius, because $D_{\rm c}$ sets the boundary between primordial planetesimals and the smaller solids in collisional equilibrium (Eq. \ref{eq:sizedist1} and \ref{eq:sizedist2}). Therefore, at a given system age, we can expect to observe 4 radial regimes within a belt's surface density distribution, arising from this critical size $D_{\rm c}$ decreasing with radius. In the innermost regions within the critical radius ($r<r_{\rm c}$), we expect the largest planetesimals to have collided and therefore all sizes to be in collisional equilibrium ($D_{\rm c}> D_{\rm max}$). Moving outwards, we then expect a region  ($r_{\rm c}<r<r_{\rm b}$) where bodies of size $D_{\rm b}$ have collided but the largest bodies have not yet ($D_{\rm b}<D_{\rm c}<D_{\rm max}$). This region corresponds to the flat lines in the right panel of Figure~\ref{fig:surface_density_model}. This region is followed by one where grains of sizes probed by our observations ($r_{\rm b}<r<r_{D_{\rm obs}}$) have collided but bodies of size $D_{\rm b}$ have not yet ($D_{\rm obs}<D_{\rm c}<D_{\rm b}$). Finally, the outermost region ($r>r_{D_{\rm obs}}$) where observable grains themselves are yet to collide ($D_{\rm c}<D_{\rm obs}$).

Here, we focus on the expected mass surface density distribution of observable grains in the innermost region ($r\leq r_{\rm c}$, where $D_{\rm c}\geq D_{\rm max}$), where the ALMA data can provide the strongest constraints. Therefore, of particular interest is the first transition in the radial dependence of the size distribution around the critical radius $r_{\rm c}$, which is defined as the location where $D_{\rm c}=D_{\rm max}$. At this location, if $D_{\rm max}>D_{\rm b}$, the size distribution from Eq. \ref{eq:sizedist1}, \ref{eq:sizedist2} and \ref{eq:sizedist3} reduces to
\begin{equation}
n(D)=n_{D_{\rm max}}\left(\frac{D}{D_{\rm max}}\right)^{2-3q_{\rm g}}\ \ \ \ \ \ \mathrm{for}\ \ D_{\rm b}<D<D_{\rm max}
\label{eq:sizedistuppernodc}
\end{equation}
and
\begin{equation}
n(D)=n_{D_{\rm max}}\left(\frac{D_{\rm b}}{D_{\rm max}}\right)^{2-3q_{\rm g}}\left(\frac{D}{D_{\rm b}}\right)^{2-3q_{\rm s}}\ \ \ \ \ \ \mathrm{for}\ \ D_{\rm min}<D<D_{\rm b}.
\label{eq:sizedistlowernodc}
\end{equation}
We can then relate the number of grains in the largest size bin, $n_{D_{\rm max}}$, to the total mass $M_{\rm tot}$ of solids in the distribution. For $D_{\rm c}=D_{\rm max}$, assuming $D_{\rm min}\ll D_{\rm b}< D_{\rm max}$ and $q_{\rm g}<2$, we have
\begin{equation}
\begin{split}
M_{\rm tot}=\frac{\pi\rho}{6}\int_{D_{\rm min}}^{D_{\rm max}}n(D)D^3dD\sim\frac{\pi\rho}{6(6-3q_{\rm g})}n_{D_{\rm max}}D_{\rm max}^4\epsilon\ \ \ \ \ \\
\mathrm{where} \ \ \ \ \ \epsilon=1+\left(\frac{6-3q_{\rm g}}{6-3q_{\rm s}}-1\right)\left(\frac{D_{\rm b}}{D_{\rm max}}\right)^{6-3q_{\rm g}}.
\label{eq:mtotndmax_dmaxgtdb}
\end{split}
\end{equation}
This allows us to express the collision rate of the largest bodies in the cascade as a function of the total mass in the size distribution, using Eq. \ref{eq:Rcolbasic} and \ref{eq:mtotndmax_dmaxgtdb}, finding
\begin{equation}
\begin{split}
R_{\rm col}(D_{\rm max})=\frac{v_{\rm rel}}{V}\frac{3(6-3q_{\rm g})}{2\rho}M_{\rm tot}\epsilon^{-1} D_{\rm max}^{3q_{\rm g}-6}\\
\int_{X_{\rm c}D_{\rm max}}^{D_{\rm max}}D_{\rm im}^{4-3q_{\rm g}}\left(1+\frac{D_{\rm max}}{D_{\rm im}}\right)^2dD_{\rm im},
\end{split}
\label{eq:rcolDmaxgtDb}
\end{equation}
where we have adopted Eq. \ref{eq:sizedistuppernodc} for the size distribution in the assumption that $X_{\rm c}>D_{\rm b}/D_{\rm max}$, i.e. the smallest impactors able to destroy an object of size $D_{\rm max}$ are larger than $D_{\rm b}$. Solving the integral leads to
\begin{equation}
\begin{split}
R_{\rm col}(D_{\rm max})=\frac{v_{\rm rel}}{V}\frac{3(6-3q_{\rm g})}{2\rho}M_{\rm tot}\epsilon^{-1} D_{\rm max}^{3q_{\rm g}-6}\\
\left[\frac{D_{\rm im}^{5-3q_{\rm g}}}{5-3q_{\rm g}}\left(1+\frac{10-6q_{\rm g}}{4-3q_{\rm g}}\frac{D_{\rm max}}{D_{\rm im}}+\frac{5-3q_{\rm g}}{3-3q_{\rm g}}\frac{D_{\rm max}^2}{D_{\rm im}^2}\right)\right]_{X_{\rm c}D_{\rm max}}^{D_{\rm max}},
\end{split}
\end{equation}
which can be significantly simplified under the assumption $X_c\ll1$ and $q_{\rm g}>5/3$, leading to
\begin{equation}
R_{\rm col}(D_{\rm max})\sim\frac{v_{\rm rel}}{V}\frac{3(2-q_{\rm g})}{2\rho(q_{\rm g}-1)}M_{\rm tot}\epsilon^{-1} D_{\rm max}^{-1}X_{\rm c}^{3-3q_{\rm g}}.
\end{equation}
We note that the $X_c\ll1$ assumption may not be a good approximation at the top of the cascade for the $Q_{\rm D}^{\star}$ law and relative velocities we considered (Fig. \ref{fig:qdstarxc}, from Eq. \ref{eq:qdstarlaw} to \ref{eq:qdstarlaw2}).

\begin{figure}
 \centering \includegraphics[trim=0.0cm 0.5cm 0.0cm 0.0cm,
    clip=true,width=1.0\columnwidth]{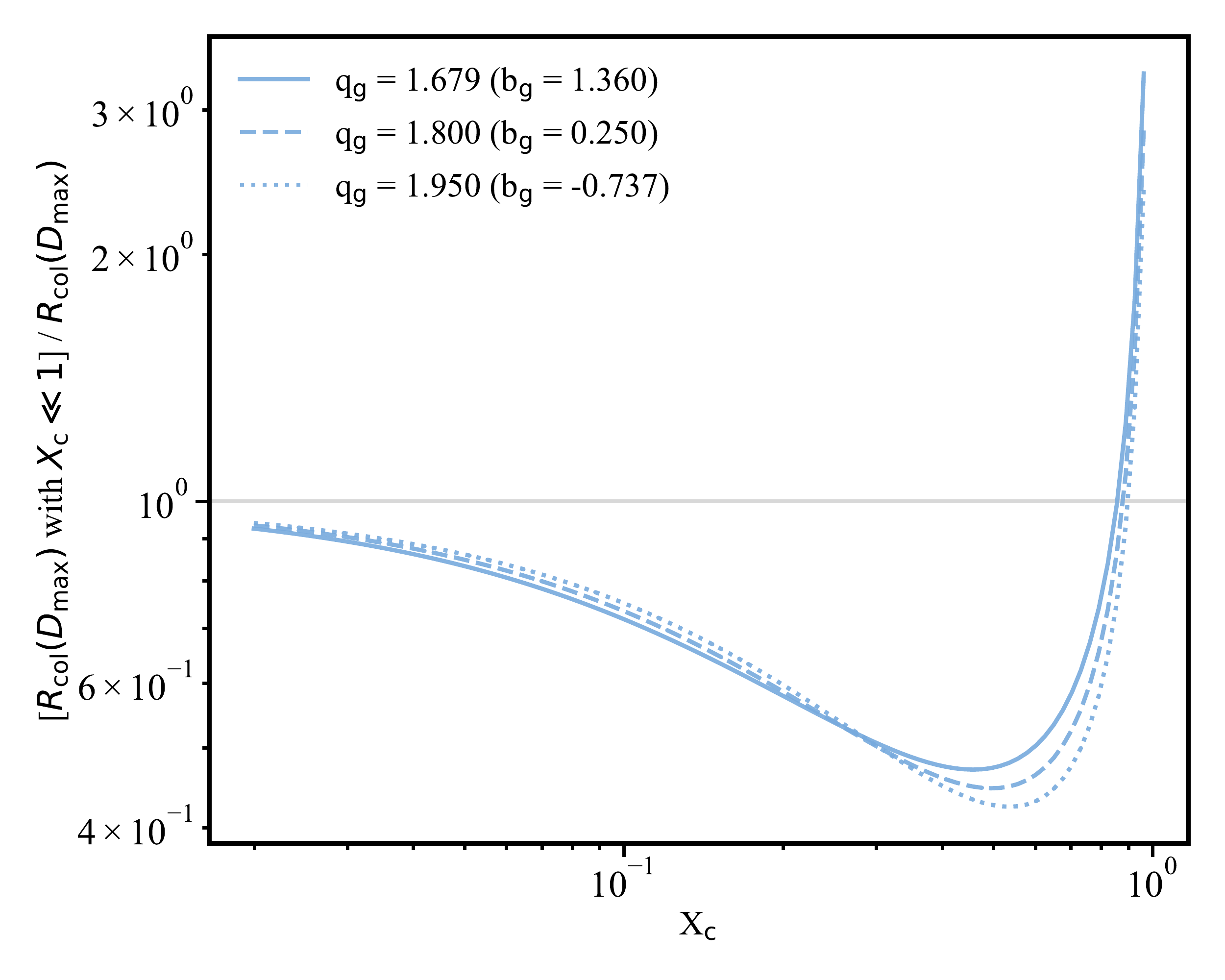}
\caption{Factor by which collision rates are over/under -estimated in the assumption $X_{\rm c}\ll1$, as a function of $X_{\rm c}$. The grey line represents a factor of 1, for which our approximation of the collision rate would be exact. For our adopted strength law, and sizes $\gtrsim10$ km, $X_{\rm c}$ can approach 1 and the collision rates therefore under/over estimated by a factor of a few. The small dependence on $q_{\rm g}$ is shown by the different line styles.} 
\label{fig:factorintegral}
\end{figure}

Fig. \ref{fig:factorintegral} shows that this approximation can underestimate the collision rates by a factor 2.4 for $X_{\rm c}$ values around 0.5, and overestimate them by a factor up to $\sim$3 for $X_{\rm c}\gtrsim0.85$. These factors have a weak dependence on $q_{\rm g}$, and for the strength law adopted, mostly affect maximum sizes $D_{\rm max}\gtrsim10$ km.  

In the next step, we 1) substitute in the definition of $X_{\rm c}$, 2) assume the volume to be that of a ring with constant vertical aspect ratio $V=4\pi r^3(dr/r)i$ (where we approximate the vertical aspect ratio encompassing the volume to be the average particle inclination $h\sim i\equiv\sqrt{\left< i^2\right>}$), 3) use the definition of $Q_{D}^{\star}$ from Eq. \ref{eq:qdstarlaw1}, 4) express $v_{\rm rel}=v_{\rm k}\sqrt{1.25\left< e^2\right>+\left< i^2\right>}=(GM_{\star})^{0.5}r^{-0.5}\sqrt{6}i$ \citep[where we assumed $\sqrt{\left< e^2\right>}=2\sqrt{\left< i^2\right>}$, e.g.][]{IdaMakino1992}, and 5) express the total mass $M_{\rm tot}$ as a surface density $\Sigma$ assuming a power law profile ($M_{\rm tot}=2\pi r^2(dr/r)\Sigma_{\rm tot}$ where $\Sigma_{\rm tot}=\Sigma_0(r/r_0)^{-\alpha}$). 
This allows us to write an expression for the collision timescale of the largest planetesimals at $r_{\rm c}$,
\begin{equation}
\begin{split}
\tau_{\rm col}(D_{\rm max})\sim\frac{2^{q_{\rm g}-1}4\rho(q_{\rm g}-1)\epsilon}{3(2-q_{\rm g})\Sigma_0 r_0^{\alpha}}D_{\rm max}^{12-6q_{\rm g}}(Q_{\rm D_b}^{\star})^{q_{\rm g}-1}D_{\rm b}^{6q_{\rm g}-11}\\
v_0^{-\frac{1}{2}q_{\rm g}+\frac{1}{2}}\left(\sqrt{6}i\right)^{-\frac{3}{2}q_{\rm g}+\frac{1}{2}}i(GM_{\star})^{-\frac{3}{4}q_{\rm g}+\frac{1}{4}}r_{\rm c}^{\left(\frac{3}{4}q_{\rm g}+\frac{3}{4}+\alpha\right)},
\end{split}
\label{eq:fulltaucol}
\end{equation}
which, in summary, applies under the assumptions $D_{\rm b}/D_{\rm max}<X_{\rm c}\ll1$, $5/3<q_{\rm g}<2$, and $D_{\rm min}\ll D_{\rm b}\ll D_{\rm max}$. This expression indicates that the collision timescale of the largest planetesimals depends on their size and bulk density ($D_{\rm max}$ and $\rho$), on the $Q_{\rm D}^{\star}$ law in the gravity regime (setting $D_{\rm b}$, $Q_{\rm D_b}^{\star}$, $q_{\rm g}$ and $v_0$), on the dynamical excitation of the planetesimals ($i$), on the stellar mass $M_{\star}$ and on the distance of the planetesimals from the star $r$, as shown in previous work \citep[e.g.][]{Lohne2008}.

\subsection{The critical radius and surface density for an undisturbed, collisionally evolving belt}
\label{sec:rcequations}

Since the condition $D_{\rm c}=D_{\rm max}$ at $r_{\rm c}$ implies $\tau_{\rm col}(D_{\rm max})=t_{\rm age}$, we can rewrite Eq. \ref{eq:fulltaucol} to find $r_{\rm c}$, obtaining
\begin{equation}
\begin{split}
r_{\rm c}\sim\left[\frac{3(2-q_{\rm g})\Sigma_0 r_0^{\alpha}}{2^{q_{\rm g}-1}4\rho(q_{\rm g}-1)\epsilon}D_{\rm max}^{6q_{\rm g}-12}(Q_{\rm D_b}^{\star})^{1-q_{\rm g}}D_{\rm b}^{11-6q_{\rm g}}v_0^{\frac{1}{2}q_{\rm g}-\frac{1}{2}} \right.\\
\left.\left(\sqrt{6}i\right)^{\frac{3}{2}q_{\rm g}-\frac{1}{2}}i^{-1}(GM_{\star})^{\frac{3}{4}q_{\rm g}-\frac{1}{4}}t_{\rm age} \right]^{\frac{1}{\frac{3}{4}q_{\rm g}+\frac{3}{4}+\alpha}}.
\end{split}
\label{eq:fullrc}
\end{equation}
We can now evaluate this equation using: $Q_{\rm D}^{\star}$ law with parameters in Table~\ref{tab:collpar}); bulk density of planetesimals of $\rho=1000$ kg m$^{-3}$ (appropriate for ice); rms inclination $i=0.025$; and taking the initial planetesimal surface density distribution to be the same as the standard Minimum Mass Solar Nebula \citep[MMSN,][]{Weidenschilling1977, Hayashi1981} with values of $\Sigma_{\rm MMSN}=270$ kg m$^{-2}$, $r_{0}=1$ au and $\alpha=1.5$ \citep{Kenyon2008} but scaled by a factor $x_{\rm MMSN}$ ($\Sigma_0=x_{\rm MMSN}\Sigma_{\rm MMSN}$). With these parameters we obtain
\begin{equation}
r_{\rm c}=55\ M_{\star}^{0.29}D_{\rm max}^{-0.53}(\epsilon^{-1} x_{\rm MMSN}t_{\rm age})^{0.28},
\label{eq:rcbetterunits}
\end{equation}
with $t_{\rm age}$ in Myr, $D_{\rm max}$ in km, $M_{\star}$ in $M_{\odot}$, and $r_{\rm c}$ in au. 

This allows us to make an analytical estimation of the radius at which the maximum size of the collisional cascade is equal to the largest body within the belt. Interior to this radius, bodies of all sizes from $D_{\rm min}$ to $D_{\rm max}$ are colliding and participating in the cascade, whereas exterior to this radius, not all sizes will have collided by the system age, with $D_{\rm c}$ lower than $D_{\rm max}$ and moving towards $D_{\rm min}$ at increasing radii. This creates a knee in the radial surface density distribution, which for planetesimals typically goes from rapidly increasing with radius interior to $r_{\rm c}$, to decreasing with a slope equal to that of the initial MMSN-like planetesimal distribution \citep[e.g.][]{Kennedy2010}\footnote{Though note that only the \textit{total} surface density distribution follows the initial planetesimal distribution outside of $r_{\rm c}$; for observable grains, the slope becomes much flatter \citep[e.g.][]{Marino2017a, Schuppler2016, Geiler2017}.}.

As well as the critical radius $r_{\rm c}$, we can estimate the surface density of observable grains ($\Sigma_{\mathrm{dust}, r=r_{\rm c}}$) at $r_{\rm c}$. The observable grains are those with a size smaller than $\sim10$ times the wavelength of interest; larger grains do not contribute significantly. For grains sizes up to this maximum observable size ($D_{\rm obs}$), we are typically in the lowest size regime of the size distribution, so Eq. \ref{eq:sizedist3} and \ref{eq:sizedistlowernodc} apply, and $n_{D_{\rm max}}$ can be once again linked to $M_{\rm tot}$ through Eq. \ref{eq:mtotndmax_dmaxgtdb}. In the assumption that $q_{\rm s}<2$ and $D_{\rm min}\ll D_{\rm obs}$, the total surface density in grains of size up to $D_{\rm obs}$ can be derived by solving the integral in Eq. \ref{eq:mtotndmax_dmaxgtdb} but with upper limit $D_{\rm obs}$ rather than $D_{\rm max}$ and using $M=2\pi r dr\Sigma$, leading to
\begin{equation}
\Sigma_\mathrm{dust}(D\leq D_{\rm obs}, r=r_{\rm c})=\frac{6-3q_{\rm g}}{6-3q_{\rm s}}\Sigma_{\rm tot}\epsilon^{-1} D_{\rm max}^{3q_{\rm g}-6}D_{\rm b}^{3q_{\rm s}-3q_{\rm g}}D_{\rm obs}^{6-3q_{\rm s}}.
\end{equation}
Expressing $\Sigma_{\rm tot}$ in terms of the surface density at $r_0$ we obtain
\begin{equation}
\Sigma_\mathrm{dust}(D\leq D_{\rm obs}, r=r_{\rm c})=\frac{6-3q_{\rm g}}{6-3q_{\rm s}}r_0^{\alpha}\Sigma_0\epsilon^{-1} r^{-\alpha}D_{\rm max}^{3q_{\rm g}-6}D_{\rm b}^{3q_{\rm s}-3q_{\rm g}}D_{\rm obs}^{6-3q_{\rm s}},
\end{equation}
which,
when inserting the same values of $q_{\rm g}, q_{\rm s}, D_{\rm b}, \Sigma_0, \alpha, r_0$, and using the same units as Eq. \ref{eq:rcbetterunits}, becomes 
\begin{equation}
\Sigma_\mathrm{dust}(D\leq D_{\rm obs}, r=r_{\rm c})=0.019\ \epsilon^{-1} x_{\rm MMSN}r_{\rm c}^{-1.5}D_{\rm max}^{-0.93}D_{\rm obs}^{0.33},
\label{eq:sigmarcbetterunits}
\end{equation}
with $\Sigma_{\mathrm{dust}, D\leq D_{\rm obs}, r=r_{\rm c}}$ in M$_{\oplus}$ au$^{-2}$ for $D_{\rm obs}$ in mm. Replacing $r_{\rm c}$ in Eq.~\ref{eq:sigmarcbetterunits} by the right hand side of Eq.~\ref{eq:rcbetterunits}, we find 
\begin{equation}
\begin{split}
    \Sigma_\mathrm{dust}( D\leq D_{\rm obs}, r=r_{\rm c}) =4.6\times10^{-5} \ \epsilon^{-0.57} x_{\rm MMSN}^{0.57}D_{\rm max}^{-0.14}D_{\rm obs}^{0.33} \\
    (t_{\rm age} M_{\star})^{-0.43}. 
\end{split}
\label{eq:sigmadustbetterunits}
\end{equation}
This expression for $\Sigma_{\mathrm{dust}}$ resembles Eq. 7 in \cite{Marino2017b}, having the same dependencies on $x_{\rm MMSN}$, $D_{\max}$, and $t$. The main difference is that Eq. \ref{eq:sigmadustbetterunits} gives a dust surface density 10 times larger (after the correction provided by \cite{Marino2019}). This difference is due to a discontinuity in the size distribution obtained using the numerical method proposed by \cite{Wyatt2011} and used in \cite{Marino2017b}. Such a discontinuity is not expected in reality and also not seen in other simulations that evolve the size distribution \citep{Lohne2008, Gaspar2012}. The numerical method tends to under-predict the dust levels by a factor 2-3 compared to the most advanced simulations of \citet{Lohne2008}. Therefore the true surface density of dust is likely a factor of  $\sim3$ smaller than equations \ref{eq:sigmarcbetterunits} and \ref{eq:sigmadustbetterunits} predict. 

Note that while Eq.~\ref{eq:sigmadustbetterunits} is valid only at $r=r_{\rm c}$, it has been shown that the surface density of dust at $r>r_{\rm c}$ is expected to be flat for a primordial surface density exponent ($-\alpha$) of -3/2, or more generally proportional to $r^{-0.6\alpha+0.9}$ \citep{Schuppler2016, Marino2017b, Geiler2017}. The flat surface density is due to two effects that balance each other out. On one hand, the surface density of solids decreases with radius. On the other hand, the size distribution at smaller radius is more collisionally eroded\footnote{For more details see \S5 in \cite{Marino2017b}.}. These two effects combined result in a dust surface density that is close to flat for a MMSN-like initial surface density. This also means that we can extrapolate $\Sigma_{\mathrm{dust}, D\leq D_{\rm obs}, r=r_{\rm c}}$ to larger radii assuming a certain $\alpha$.

\begin{figure*}
\vspace{-0mm}
 \centering
 \hspace{-5mm}
   \includegraphics*[scale=0.48]{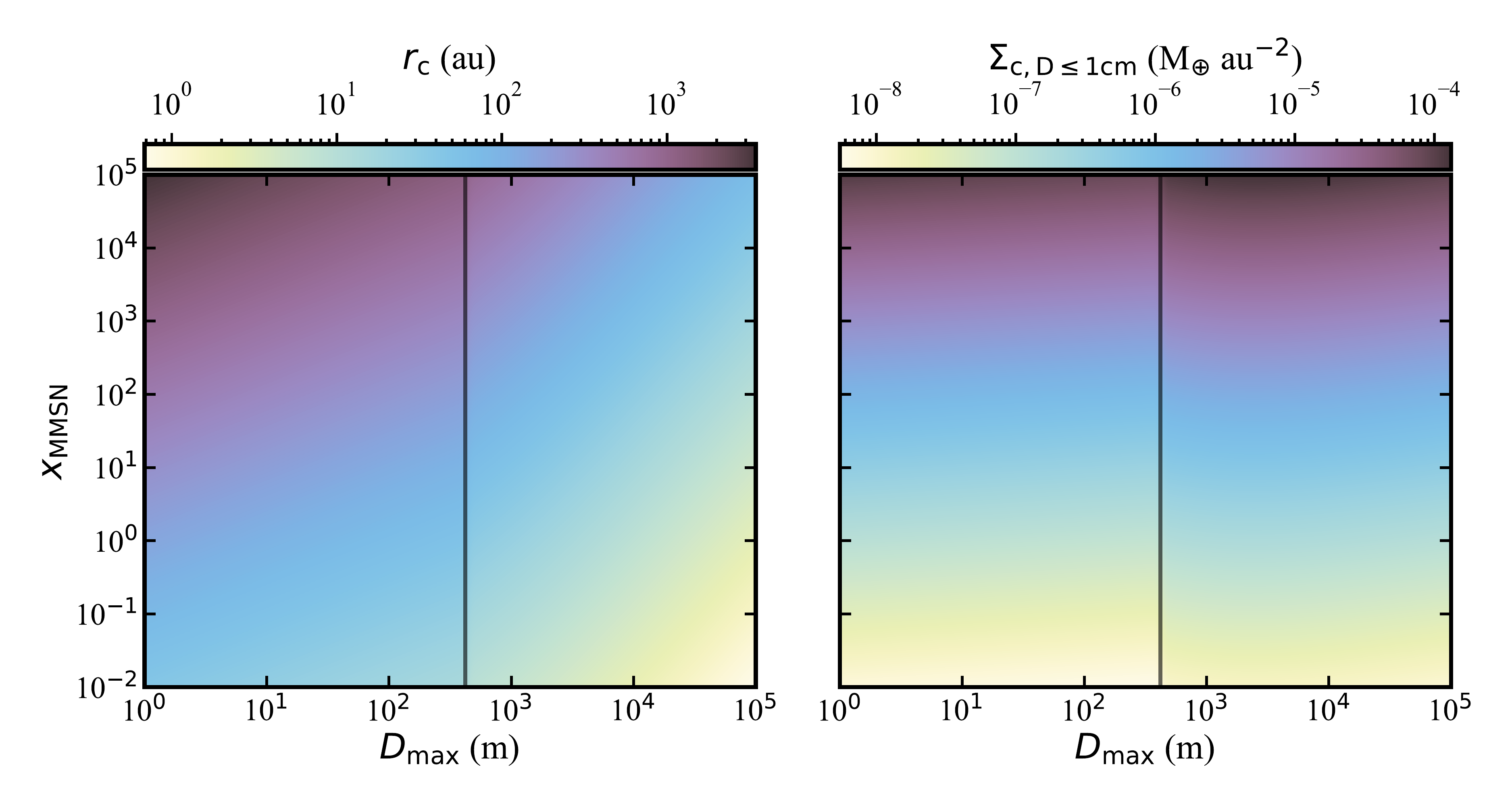}
\vspace{-8mm}
\caption{The colour scale shows the dependence of the critical radius $r_{\rm c}$ (left panel) and the surface density of observable grains $\Sigma_{\rm c, D\leq1cm}$ at that radius (right panel) on the maximum size of planetesimals in the size distribution ($D_{\rm max}$) and on the initial planetesimal disc mass ($x_{\rm MMSN}$, as a multiple of the MMSN). The vertical line represents the size boundary $D_{\rm b}$ between the strength and gravity regime of our $Q_{\rm D}^{\star}$ law, at which there is a discontinuity in the colour scale due to our analytical approximation (see text for details).} 
\label{fig:rc_sigmac_vsDmax_vsxmmsn}
\end{figure*}

Together with Eq. \ref{eq:rcbetterunits}, Eq. \ref{eq:sigmarcbetterunits} implies that if we know the age of a system and can accurately measure both the critical radius and the surface density of grains at that radius,  
rearranging Eq. \ref{eq:rcbetterunits} and \ref{eq:sigmarcbetterunits} and keeping the same units, we can explicitly derive $D_{\rm max}$ and $x_{\rm MMSN}$ from observables $r_{\rm c}$ and $\Sigma_{\rm c}\equiv\Sigma_{\mathrm{dust}, D<D_{\rm obs}, r=r_{\rm c}}$, obtaining 
\begin{eqnarray}
D_{\rm max}=2.6\times10^{8}\ M_{\star}^{1.09}(t_{\rm age}\Sigma_{\rm c})^{1.07}r_{\rm c}^{-2.16}D_{\rm obs}^{-0.35} \label{eq:dmaxgbetterunits}\\
x_{\rm MMSN}=3.8\times10^{9}\ \epsilon M_{\star}^{1.02}t_{\rm age}\Sigma_{\rm c}^{2}r_{\rm c}^{-0.52}D_{\rm obs}^{-0.65}. \label{eq:xmmsngbetterunits}
\end{eqnarray}
The above assumes we are in the regime where $D_{\rm max}>D_{\rm b}$ and $X_{\rm c}D_{\rm max}>D_{\rm b}$. If instead the entire size distribution is in the strength regime of the $Q_D^{\star}$ law (i.e. $D_{\rm max}<D_{\rm b}$), we have 
\begin{eqnarray}
r_{\rm c}=81\ M_{\star}^{0.32}D_{\rm max}^{-0.18}(x_{\rm MMSN}t_{\rm age})^{0.27},\\
\Sigma_{\rm c} = 1.5\times10^{-5} M_{\star}^{-0.48}t_{\rm age}^{-0.41}x_{\rm MMSN}^{0.59}D_{\rm obs}^{0.33} D_{\rm max}^{-0.06}. \label{eq:sigmacsbetterunits} \\
D_{\rm max}=2.7\times10^{27}\ M_{\star}^{3.58}(t_{\rm age}\Sigma_{\rm c})^{3.07}r_{\rm c}^{-6.65}D_{\rm obs}^{-1}  \label{eq:dmaxsbetterunits} \\
x_{\rm MMSN}=8.0\times10^{10}\ M_{\star}^{1.17}t_{\rm age}\Sigma_{\rm c}^{2}r_{\rm c}^{-0.67}D_{\rm obs}^{-0.65}. \label{eq:xmmsnsbetterunits}
\end{eqnarray}
When comparing Equation \ref{eq:sigmacsbetterunits} with the numerical model from \cite{Marino2017b}, we find a better match than for $D_{\max}>D_{\rm b}$ with a difference of a factor of $\sim3$.

Fig. \ref{fig:rc_sigmac_vsDmax_vsxmmsn} visualizes the dependence of $r_{\rm c}$ and $\Sigma_{c}$ for grains up to 1~cm on $D_{\rm max}$ and $x_{\rm MMSN}$ for a 100 Myr-old system around a Sun-like 1.0 M$_{\odot}$ star. While $r_{\rm c}$ has a clear dependence on $D_{\rm max}$, $x_{\rm MMSN}$ and $t_{\rm age}$, $\Sigma_{\rm c}$ only has weak dependence on $D_{\rm max}$, and so is mostly sensitive to the scaling of the total initial planetesimal mass, $x_{\rm MMSN}$ and $t_{\rm age}$. As noted by \citet{Marino2017a}, this is due to a balance between i) higher $D_{\rm max}$ values producing lower dust masses for the same total mass $M_{\rm tot}$ in the size distribution, and ii) higher $D_{\rm max}$ values leading to longer collision timescales at the top of the cascade and therefore slower collisional evolution and higher dust masses for a given system age. As a belt evolves collisionally (larger $t_{\rm age}$), $r_{\rm c}$ (i.e. the $D_{\rm c}=D_{\rm max}$ point) moves outwards and $\Sigma_{\rm c}$ decreases as a result, due to the lower initial surface densities at larger radii (for a MMSN-like planetesimal disc). More massive host stars $M_{\star}$ produce larger collision velocities and faster collisional processing; therefore, within a given age and initial surface density, $r_{\rm c}$ will have moved further out and the dust surface densities $\Sigma_{\rm c}$ will be lower around more massive stars.


Finally, note that using the expressions above cause a discontinuity at $D_{\rm max}=D_{\rm b}$ (black vertical line in Fig. \ref{fig:rc_sigmac_vsDmax_vsxmmsn}). This is because just above $D_{\rm b}$ we have $X_{\rm c}D_{\rm max}<D_{\rm b}<D_{\rm max}$, so it is not possible to simplify the collision rate integral in Eq. \ref{eq:rcolDmaxgtDb} as done above. This unfortunately implies that we cannot extract the $r$ dependence from the collision rate $R_{\rm col}$ (and later timescale $\tau_{\rm col}$) analytically, but only establish a limit to $D_{\rm max}$ in the range $[D_{\rm b},D_{\rm b}/X_{\rm c}(D_{\rm max})]$.

\subsection{The radial slope of the surface density interior to the critical radius}
\label{sec:radialslope}
Going one step further, we can establish the expected slope interior to the critical radius ($r\ll r_{\rm c}$) of a collisionally evolving, undisturbed planetesimal belt.
The procedure is the same, starting from Eq. \ref{eq:sizedistlowernodc}, but this time substituting $n_{D_{\rm max}}$ with the collisionally evolved density $n_{D_{\rm max}} \tau_{\rm col}(D_{\rm max})/t_{\rm age}$  in Eq. \ref{eq:mtotndmax_dmaxgtdb}. This is because at radii smaller than $r_{\rm c}$, all sizes participate in the cascade just like at $r=r_{\rm c}$, but because the largest bodies are colliding, the overall mass of the cascade is decreasing and scaled down by a factor $(1+t_{\rm age}/\tau_{\rm col}(D_{\rm max}))^{-1}$ at $t=t_{\rm age}$ compared to the initial mass. Since we are interested in $r\ll r_{\rm c}$, we take the approximation $t_{\rm age}\gg\tau_{\rm col}$ to obtain 
\begin{equation}
\Sigma_{\mathrm{tot},t_{\rm age}, r\ll r_{\rm c}}=\Sigma_{\mathrm{tot}, t_{0}, r= r_{\rm c}}\tau_{\rm col}(D_{\rm max})t_{\rm age}^{-1}.
\end{equation}
Inserting $\tau_{\rm col}(D_{\rm max})$ from Eq. \ref{eq:fulltaucol}, we derive
\begin{equation}
\begin{split}
\Sigma_\mathrm{dust}( D\leq D_{\rm obs}, r\ll r_{\rm c})=\frac{2^{q_{\rm g}+1}(q_{\rm g}-1)}{3(2-q_{\rm s})}\rho D_{\rm max}^{6-3q_{\rm g}}D_{\rm obs}^{6-3q_{\rm s}}(Q_{\rm D_b}^{\star})^{q_{\rm g}-1}\\
D_{\rm b}^{3q_{\rm g}+3q_{\rm s}-11}v_0^{-\frac{1}{2}q_{\rm g}+\frac{1}{2}}\left(\sqrt{6}i\right)^{-\frac{3}{2}q_{\rm g}+\frac{1}{2}}i(GM_{\star})^{-\frac{3}{4}q_{\rm g}+\frac{1}{4}}r^{\frac{3}{4}q_{\rm g}+\frac{3}{4}}t_{\rm age}^{-1},
\end{split}
\end{equation}
which is independent of the initial planetesimal surface density distribution $\Sigma_{\rm tot}$ and its parameters $\Sigma_0, \alpha$, and $r_0$. We can simplify this equation using the same values of $q_{\rm g}, q_{\rm s}, D_{\rm b}, \rho$, $Q_{\rm D_b}^{\star}, i$, and $v_0$ and units used so far and summarised in Table~\ref{tab:collpar}, finding 
\begin{equation}
    \Sigma_{\rm dust}( D\leq D_{\rm obs}, r\ll r_{\rm c})=5.4\times10^{-7} D_{\max}^{0.93} D_{\rm obs}^{0.33} M_{\star}^{-1.02} r^{2.02} t_{\rm age}^{-1},
\end{equation}
where $\Sigma_{\rm dust}$ is in units of $M_{\oplus}$~au$^{-2}$, $D_{\max}$ in km, $D_{\rm obs}$ in mm, $M_{\star}$ in $M_{\odot}$, $r$ in au, and $t_{\rm age}$ in Myr.

We find that the slope $\gamma$ of the surface density of grains in this regime, i.e. interior to the belt's critical radius, should be positive (surface density increasing with radius) and equal to $\gamma=0.75q_{\rm g}+0.75=2.02$. Therefore, the inner surface density slope for an undisturbed, collisionally evolving planetesimal belt is solely determined by the slope of the size distribution at the distribution's upper end. This comes from the slope of the $Q_{\rm D}^{\star}$ law in the gravity regime if $D_{\rm max}>D_{\rm b}$,    or in the strength regime if $D_{\rm max}<D_{\rm b}$. Note that the slope $\gamma$ is slightly different from the value of $7/3$ obtained in more simple analytical models due to considering a $Q^{\star}_{\rm D}$ independent of size and velocity \citep{Kennedy2010}. Nevertheless, for typical values of $q_{\rm g}$ \citep[e.g. 1.9, ][]{BenzAsphaug1999}, they differ by less than 10\%.




\section{Inner edge constraints from the data} \label{sec:dataconstraints}

In this section, we aim to constrain the inner surface density slope of several wide debris discs that have been well resolved with ALMA. Determining the slope will allow us to assess whether the inner edge is consistent with being set by collisional evolution alone, or instead, the disc was truncated at the inner edge, for example, by a planet. The inner slope is retrieved by fitting a parametric model directly to the ALMA visibilities. In order to choose the right parametric model, we first use the {\sc frankenstein} {\sc Python} package \citep[hereinafter referred to as \textsc{frank};][]{Jennings2020} that reconstructs the intensity radial profiles in a non-parametric manner and achieves higher resolutions than typical clean images. The \textsc{frank} recovered profiles allow us to have a clearer idea of the different features in each system that need to be fit and thus we can make a more informed decision in choosing a parametric model to best fit each target. Furthermore, they also allow for a more consistent approach in how radial profiles are determined. Below we describe the chosen targets, the \textsc{frank} profiles, and the results after fitting the parametric models.

\subsection{Targets}

\begin{table*}
\centering
\caption{General information on the 7 systems studied: HD~92945, HD~107146, HR~8799, q$^{1}$~Eri, AU~Mic, 49~Ceti, and HD~206893. The 6$^{\rm th}$, 7$^{\rm th}$ and 8$^{\rm th}$ columns show the discs' fractional luminosities (this work), inclinations and position angles. Sources: (1) \protect\cite{Marino2021}, (2) \protect\cite{Lovell2021}, (3) \protect\cite{Hughes2017}, (4) \protect\cite{Torres2006}, (5) \protect\cite{Harlan1970}, (6) \protect\cite{Gray1999}, (7) \protect\cite{Williams2004}, (8) \protect\cite{Plavchan2009}, (9) \protect\cite{Bell2015}, (10) \protect\cite{Marmier2013}, (11) \protect\cite{Roberge2013}, (12) \protect\cite{Zuckerman2012}, (13) \protect\cite{Gray2006}, (14) \protect\cite{Watson2011}, (15) \protect\cite{Sepulveda2022}, (16) \protect\cite{Mamajek2014}, (17) \protect\cite{Plavchan2020}, (18) \protect\cite{Hinkley2022},  (19) \protect\cite{Gaia2021}, (20) \protect\cite{Marois2010},  (21) \protect\citet{Zurlo2022}, (22) \protect\cite{Wittrock2022}.}
\label{tab:basicdatasystems}
\setlength\tabcolsep{2.0pt}
\begin{tabular}{lccccccccc}
\hline
\hline
System &  Distance  &  Spectral Type & Age  & Stellar mass & $f_{\rm dust}$ & Inclination  & PA  & Planet semi-major axis & Planet mass    \\ 	
& [pc] &  & [Myr]  & [$M_\odot$] &  & [deg] & [deg]  & [au] & [$M_{\rm Jup}$] \\
\hline
HD~92945  & 21.5 (19)  & K1V (4) & $294\pm23$ (17) & $0.86\pm0.01$ (8) & $7\times10^{-4}$ & $65.4 (1)$ & $100 (1)$  \\
HD~107146 & 27.5 (19)  & G2V (5) & $80-200$ (7)   & 1.09 (14)    &    $10^{-3}$  & $19.9$ (1) & $153 (1)$  \\
HR~8799   & 41.3   (19) & A5 (6) & $42^{+6}_{-4} (9)$ & $1.43^{+0.06}_{-0.07}$ (15) & $3\times10^{-4}$ & $31.2$ (1) & $52.0$ (1) & 16, 27, 41, 71 (20,21) & 8, 9, 8, 6 (20,21) \\
q$^{1}$~Eri & 17.4  (19)  &  F9V (10) & $(1.4\pm0.9)\times10^3 $ (10)  &$1.11\pm0.02$ (10) & $3\times10^{-4}$ & $78.6$ (2) & 57.0 (2) & 2 (10) &  1 (10)   \\ 
AU~Mic  & 9.7 (19) & M1V (4) & $22\pm3$ (16) &$0.50\pm0.03$  (17) & $4\times10^{-4}$ & 88.2 (1) & 128.5 (1) & 0.065, 0.11 (22) &  0.05, 0.007-0.079 (22) \\
49~Ceti  & 57.2 (19) & A1V (11) & 40 (12) & 2.1 (3) & $7\times10^{-4}$ & 79.1 (3) & 107.4 (3)  \\
HD~206893  & 40.8 (19) & F5V (13)  & 170 (18) & $1.32^{+0.07}_{-0.06}$ (17) & $3\times10^{-4}$ & 40.0 (1) & 61.7 (1) & 3.5, 9.7 (18) & 12, 27 (18) \\
\hline
\end{tabular}
\end{table*}

We focus on systems with wide exoKuiper belts that have been observed with ALMA at a sufficient resolution (those with a radial span that has been resolved with ${\gtrsim}5$ beams across) and sensitivity to characterize their inner slope (signal-to-noise ratios larger than 10 near the inner edge in the azimuthally averaged radial profiles), and that are not very asymmetric or too large as to require multiple pointings with ALMA \citep[e.g. $\beta$~Pic][]{Matra2019b}. We identify HD~92945, HD~107146, HD~206893, HR~8799, q$^{1}$~Eri, AU~Mic and 49~Ceti as the best systems to do this, all located at distances ranging from 10 to 60 pc, having wide discs, and observed with ALMA at sufficient resolution and sensitivity to constrain the inner edge shape. 

We use both 0.88mm (band 7) and 1.33mm (band 6) published data of HD~107146\footnote{We do not include data from \cite{Ricci2015a} in our analysis due to its lower resolution and sensitivity.}, HD~206893 and q$^{1}$~Eri \citep{Marino2018b, Marino2019, Marino2020hd206, Nederlander2021, Lovell2021}, 49~Ceti's 0.61~mm (band 8) data \citep{Higuchi2019}\footnote{We do not use the published band 6 data that has a much poorer resolution \citep{Hughes2017}.}, and the band 7 data of HR~8799\footnote{We do not use the published band 6 data that has a much poorer resolution \citep{Booth2016}.}, HD~92945 and AU~Mic\footnote{New 0.45mm (band 9) data were published while writing this paper, but involved multiple pointings and thus were omitted them from our analysis which cannot account for that.} \citep{Marino2019, Faramaz2021, Daley2019}. In addition to the published data on HD~107146 \citep{Marino2018b, Marino2021}, we include a new data set with a higher resolution (0.2~arcsec = 5~au) from an unfinished cycle 7 program (2019.1.00189.S). This new data set is described in Appendix \ref{app:data}. 

Finally, for HR~8799 and HD~107146 we subtract emission from a background galaxy prior to any analysis using the best parameters found in \cite{Marino2021}. Basic information of the targets in this investigation can be found in Table \ref{tab:basicdatasystems}.

\subsection{Deconvolved Profiles}
\label{sec:deconvolvedprofiles}

Prior to fitting a parametric model of the intensity radial profiles to the data, we fit them in a non-parametric way using \textsc{frank} to avoid introducing biases from the start. \textsc{frank} has one great advantage over standard imaging \citep[e.g. with \textsc{tclean} in \textsc{CASA},][]{casa}, which is that it provides a significantly better resolution than clean images. However, there are some assumptions that it makes and some parameters that must be adjusted for it to provide appropriate radial profiles. The first assumption that \textsc{frank} makes is that the discs are axisymmetric as it only fits the real component of the deprojected visibilities. This is mostly a valid assumption for these targets, however, \cite{Lovell2021} and \cite{Marino2019} find that the discs around q$^{1}$~Eri and HD~92945 show some minor asymmetries that could bias our results for those targets.


\begin{figure}
    \centering
    \includegraphics[trim=2.0cm 0.0cm 3.0cm 0.0cm, clip=true, width=1\columnwidth]{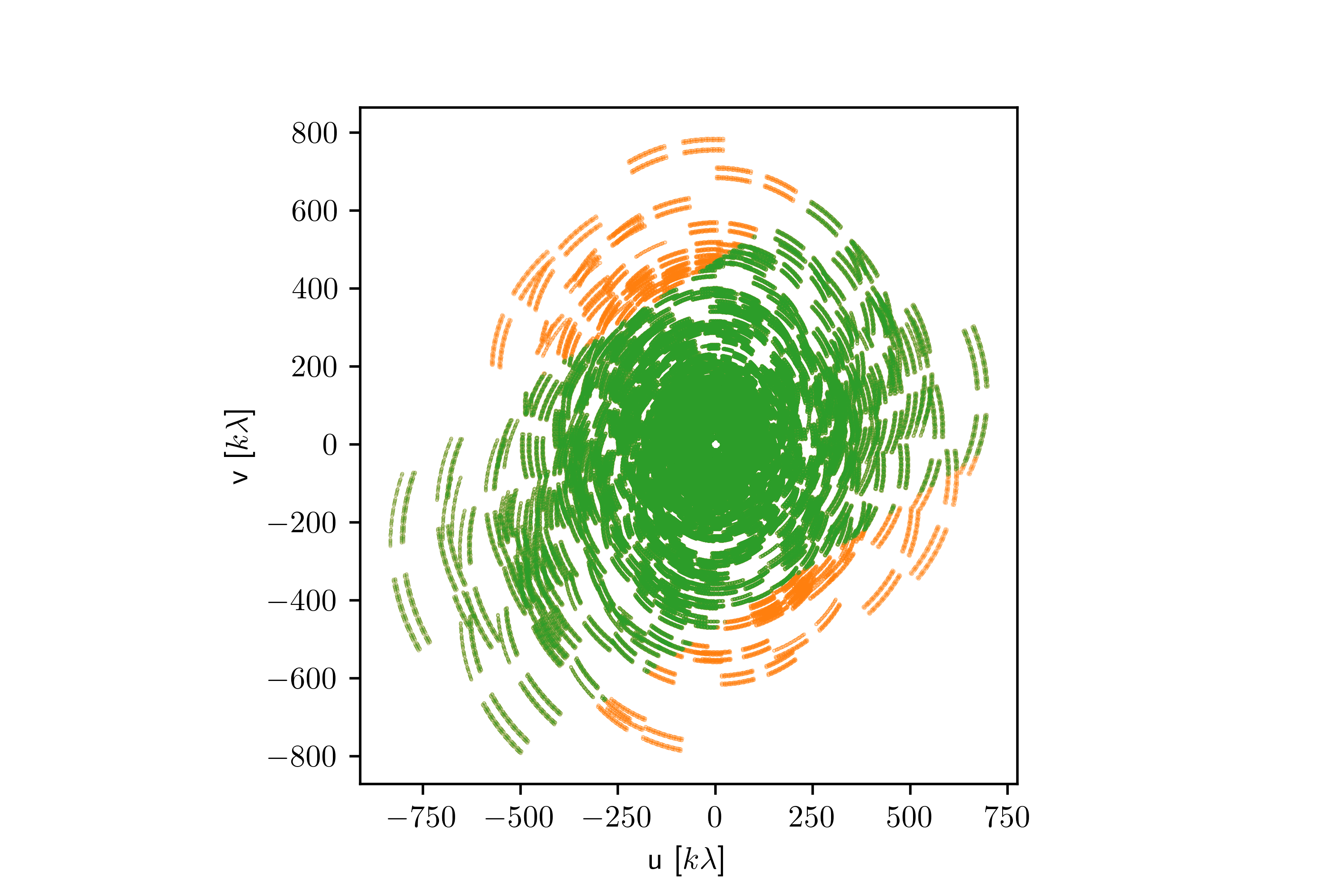}
    \caption{UV coordinates of q$^{1}$~Eri band 7 observations. The points in orange represent the baselines that are most affected by the vertical thickness of the disc and thus are removed from our analysis.}
    \label{fig:VerticalMask}
\end{figure}

\begin{figure*}
    \includegraphics[width=0.33\textwidth]{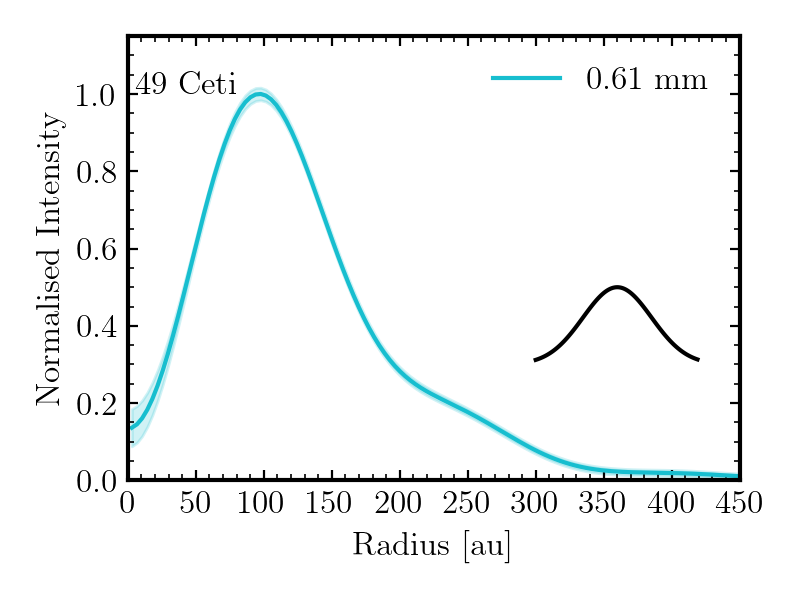}
    \includegraphics[width=0.33\textwidth]{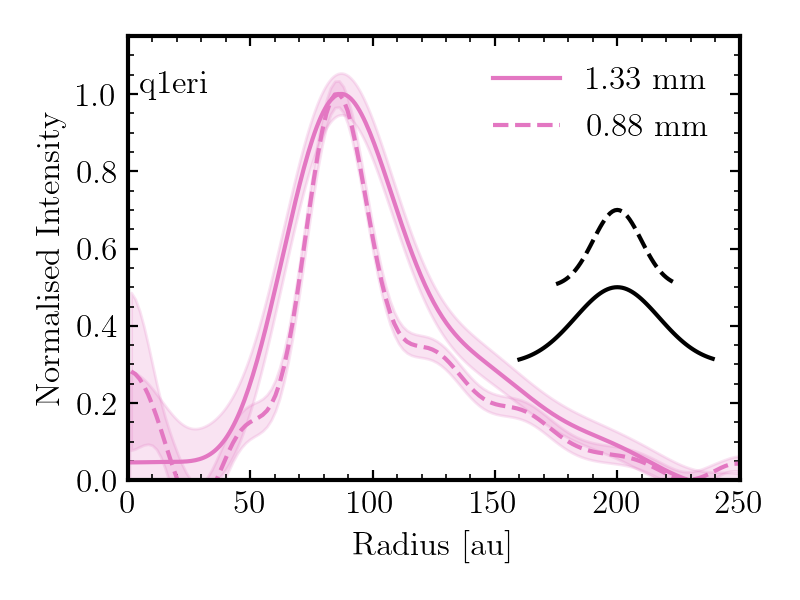}
    \includegraphics[width=0.33\textwidth]{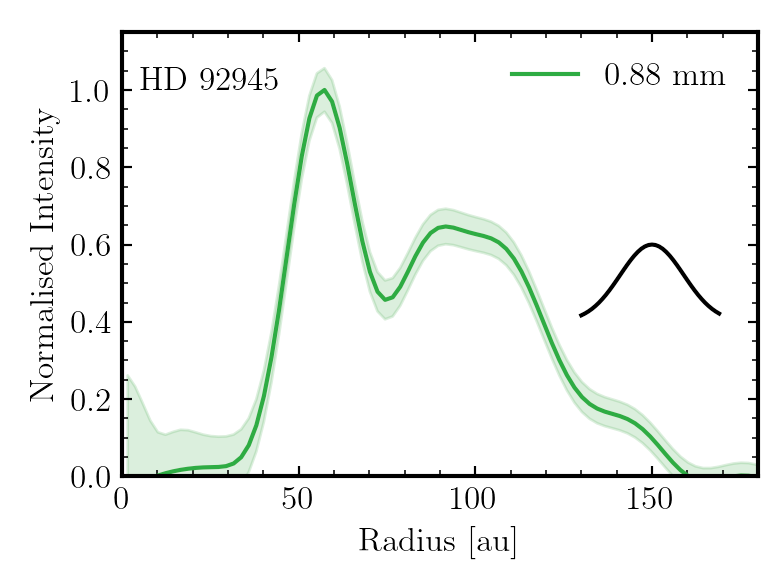}
    \includegraphics[width=0.33\textwidth]{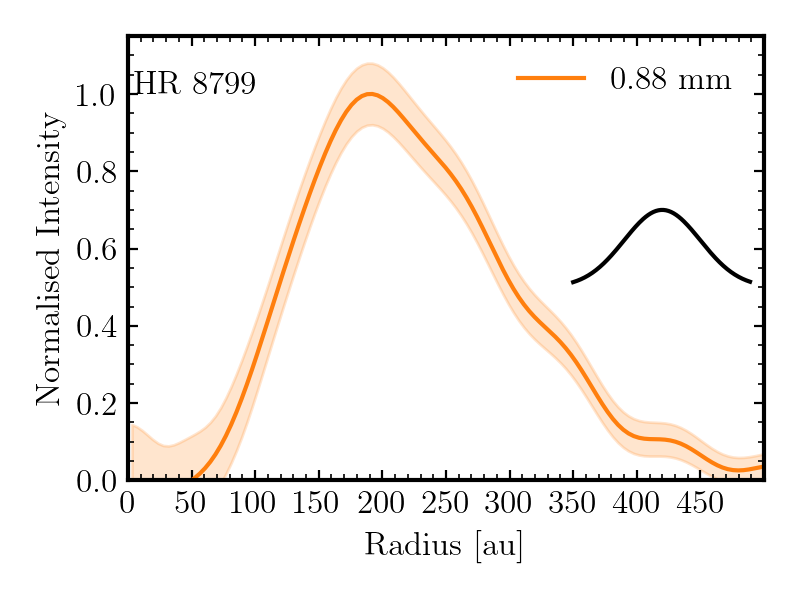}
    \includegraphics[width=0.33\textwidth]{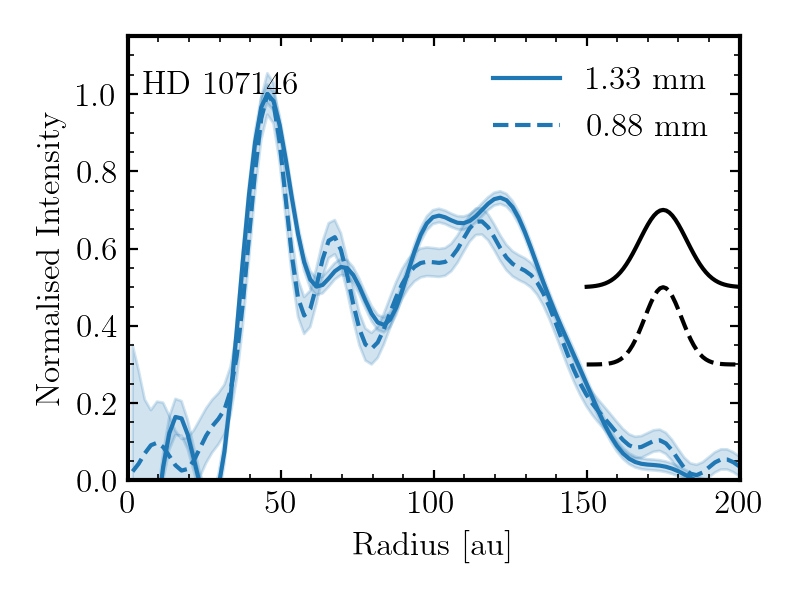}
    \includegraphics[width=0.33\textwidth]{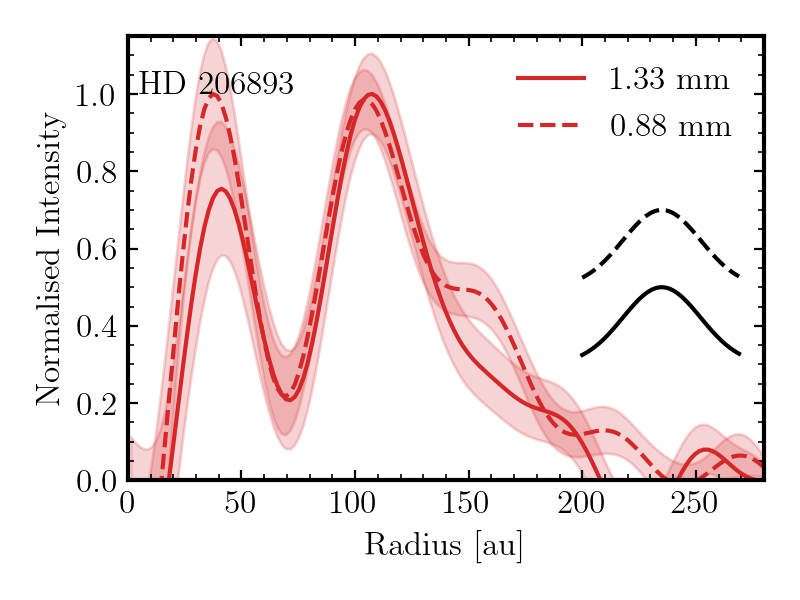}
    \includegraphics[width=0.33\textwidth]{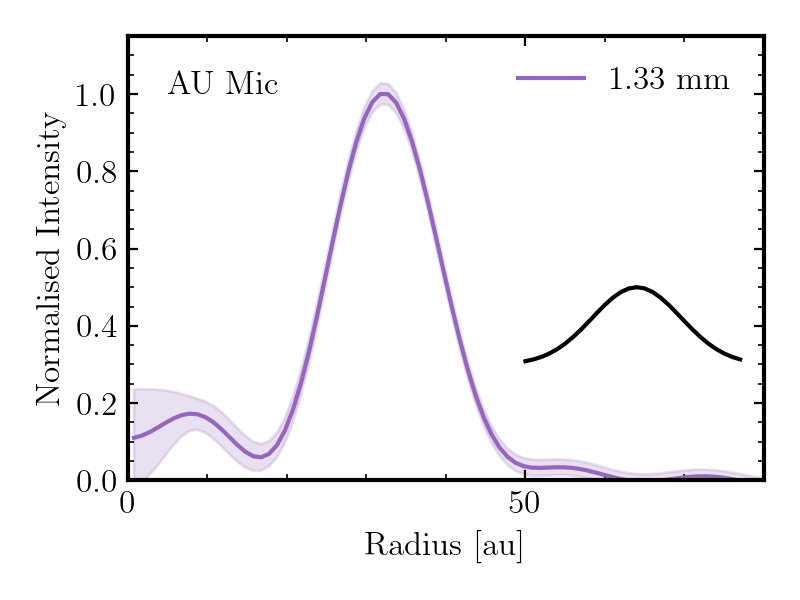}

\caption{Radial Profiles of the systems investigated, fitted using \textsc{frank}. The shaded regions represent the $1\sigma$ uncertainties derived from  \textsc{frank}. All surface brightness profiles are normalized to the peak intensity. The black lines represent the resolution of the \textsc{frank} profiles, obtained by identifying the baselines beyond which the power spectrum obtained by \textsc{frank} is damped.}
\label{fig:FrankProfiles}
\end{figure*}

The second assumption made by \textsc{frank} is that the discs are vertically flat, which has been shown to not be the case for debris discs \citep{Matra2019b, Daley2019}. In particular, previous analysis of q$^{1}$~Eri and HD~92945 found both discs to be marginally resolved with  vertical aspect ratios of ${\sim}0.05$ \citep{Marino2019, Lovell2021}. In order to account for this,  data from uv points that could be heavily affected by the vertical thickness of the disc are removed from the analysis. These points can be roughly identified as those where the uv coordinate parallel to the minor axis of the disc ($v'$) is large enough to resolve the projected vertical thickness or full width half maximum (FWHM) of the disc. This maximum baseline is estimated as 
\begin{equation}
    v'=\left(2.355\frac{r_{\rm belt}\ h}{d}\sin(i)c \right)^{-1},       \label{eq:verticalsolved}
\end{equation}
where $r_{\rm belt}$ is the central radius of the disc, $h$ is the vertical aspect ratio (a quantity that was assumed to be $0.05$ in agreement with the typical values derived for these and other systems), $d$ is the distance to the disc from Earth, $i$ is the inclination of the disc with respect to being face-on. The term inside the brackets is the projected vertical FWHM at the disc central radius. The factor 2.355 is to convert from the vertical standard deviation to a FWHM. The value for $v'$ was then used to filter the data before using \textsc{frank}, removing the data that would be most affected by the vertical thickness of the disc. 

Figure \ref{fig:VerticalMask} shows as an example the baselines that get removed through this method (orange) for q$^{1}$~Eri's band 7 observations, representing 4\% of all visibilities. Only 0.7\% of the complementary band 6 data was removed. For HD~92945, only $0.05\%$ of the 12m array data got removed and none of the ACA data.  HD~206893 had 1.52\% of band 7 and 0\% of band 6 data removed. AU~Mic, which is edge-on, had 2.24\% of its 12m data removed. Finally, none of the 12m and ACA data of HD~107146, HR~8799 and 49 Ceti was removed. These percentages vary greatly from system to system as they have different sizes, distances, inclinations, and each was observed at different resolutions. Whilst these may not seem like high percentages of removal, they could significantly distort the \textsc{frank} results and thus their removal was important, particularly for the more edge-on discs. Note that there might still be some minor vertical information in the remaining visibilities (e.g. if the signal-to-noise is high or $h$ is higher than assumed), but its effect on the deprojected visibilities and recovered profile should be minor (Terrill et al. submitted). After this process of removing data affected by vertical thickness, the flux of each of the stars was removed in the visibility space\footnote{The stars and discs were well centred at the phase centre of these observations and thus we subtract the stellar flux simply as a constant from the real component of the visibilities.} prior to running \textsc{frank} so that it would not bias the retrieved radial profiles \citep{Jennings2022}. 


Before applying \textsc{frank} to the discs, some parameters must be determined for the fits to work properly. These are known in \textsc{frank} as  $\alpha$ and $w_{\rm smooth}$. $\alpha$ is a parameter that roughly defines the maximum baseline to which \textsc{frank} will try to fit the data, acting as a signal-to-noise threshold, a higher value imposing a stricter signal-to-noise threshold. $w_{\rm smooth}$ is a spectral smoothness parameter, with a higher value more strongly smoothing the power spectrum. For more details on these parameters refer to \cite{Jennings2020}, and the \textsc{frank} \href{https://discsim.github.io/frank/index.html}{documentation}. In order to determine an appropriate value for these two parameters, we run \textsc{frank} for each system several times whilst varying the values of the parameters. The most appropriate ones were determined by visually inspecting the recovered profiles and minimising the number of oscillatory artifacts, whilst still trying to recover sharp features along the entire radius of the discs. The values tested for $\alpha$ were: [1.001, 1.01, 1.1], and for $w_{\rm smooth}$ the values tested were [$10^{-4}$, $10^{-3}$, $10^{-2}$, $10^{-1}$]. The final values for these parameters decided for each disc and band had mostly $\alpha=1.01$, with one exception of HD~107146 band 6 having an $\alpha=1.1$. The $w_{\rm smooth}$ chosen parameter ranged from $10^{-2}$ to $10^{-4}$ depending on the disc. The main features found in each of the profiles using \textsc{frank} were not particularly sensitive to the chosen parameters and were visible in most tested parameters and thus they can be considered robust. 

The radial profiles using \textsc{frank} for the seven targets can be seen in Figure \ref{fig:FrankProfiles}. The differences between bands/wavelengths for q$^1$~Eri, HD~107146, and HD~206893 are due to the differing resolutions for the data sets, making features appear smoother in one band compared to the other. For q$^1$~Eri, the profile was found to be a narrow peak at ${\sim}90$~au with a long extension out to 200~au. The profiles found by \textsc{frank} for q$^1$~Eri are consistent with previous analysis using clean images and parametric modelling by \cite{Lovell2021}. The profile resolved for HD~92945 shows a gap centred at approximately 75au, consistent with \cite{Marino2019}. The profile resolved for HR~8799 shows a broad peak at ${\sim}200$~au with smooth inner and outer edges, which is consistent with \cite{Faramaz2021}. For AU~Mic, \textsc{frank} finds a broad peak at $\sim35$~au and a tentative gap in the disc at $\sim15$~au, also found in parametric fits to the data \citep{Daley2019, Marino2021}. For 49~Ceti, we find a wide peak at $\sim100$~au and a slowly decreasing outer edge, which agrees with previous studies \cite{Hughes2017,Pawellek2019}. Finally, for HD~206893 \textsc{frank}, finds a steep inner edge with two peaks at $\sim40$~au and $\sim120$~au and a deep gap in between centred at $\sim75$~au \citep[consistent with][]{Marino2020hd206, Nederlander2021}.

The main disc for which there is a significant difference between this analysis and previous findings is HD~107146 \citep{Ricci2015a, Marino2018b, Marino2021}. \citeauthor{Marino2018b} found one wide and shallow gap in the radial profile. However, using \textsc{frank} we find that this wide gap is split into two narrow ones which previous clean images did not resolve due to their poorer resolution. This double gap structure is found in both the band 6 and 7 data of HD~107146 shown in Figure~\ref{fig:FrankProfiles} (with a higher significance in the band 7 data due to its higher resolution). Therefore, we consider this to be a true feature rather than an artifact produced by \textsc{frank}. Moreover, this feature is also revealed in the radial profile extracted from new higher resolution clean images in band 7 presented in Appendix~\ref{app:data}. These narrower gaps could be consistent with the scenario proposed by \cite{Marino2018b} where two 10~$M_{\oplus}$ planets at separations between $50-90$~au could carve two independent gaps, which at low resolution appeared as one half-empty wide gap. Note that previous work that fit parametric models to the data did not try fitting a double gap model, leaving this feature undiscovered. This highlights the importance of using \textsc{frank} first to visualize the radial features of the disc. This system will be observed by JWST in 2023 with MIRI at 15~$\mu$m in coronagraphic mode to search for companions above a 0.2~$M_{\rm Jup}$ beyond 20~au \citep{Marino2021jwst}. Such observations, combined with the double gap structure, will allow for a much clearer interpretation. Therefore, we defer the interpretation and discussion of this feature until the JWST data becomes available.

The main disadvantage of \textsc{frank} for this investigation is that it does not directly provide estimates for the slope of the inner edge, which is the aim of this investigation. Whilst this could be measured from the recovered radial profiles, such measurements would be affected by non-trivial systematic effects such as \textsc{frank}'s non-Gaussian PSF that are hard to account for and thus could bias our results. Instead, the profiles achieved with \textsc{frank} can be used to decide which parametric models are most appropriate in order to constrain the steepness of the inner edge. 

\subsection{Parametric Fits to the Data/Visibilities}
\label{sec:fits}

As we are only interested in the radial profiles of these discs, we can azimuthally average the visibilities and deproject them assuming inclinations and position angles derived in previous studies, found in Table \ref{tab:basicdatasystems}. As in \S\ref{sec:deconvolvedprofiles}, before fitting the data we removed the data points that could be affected by the vertical thickness of the disc. In order to speed up the fitting process of millions of $u-v$ points, we binned the visibilities as a function of their deprojected uv distance. The visibilities were binned with a bin width set to be 5\% of the smallest uv data point in each individual data set. We find this width is small enough to not lose the details in the visibility profile, and large enough to reduce significantly the number of data points being fit. Within each bin, we determined the uncertainty as the standard deviation divided by the square root of the number of data points. Note that we only consider the real component of the visibilities as it is assumed the discs are axisymmetric and thus their imaginary component is zero. Previous analysis by \cite{Marino2019} and \cite{Lovell2021} showed some significant, but minor asymmetries in the discs around HD~92945 and q$^{1}$~Eri. Those asymmetries are mostly located beyond the disc inner edge, and thus we consider they should not affect our results and conclusions significantly. Moreover, asymmetries will tend to smooth the azimuthally averaged profile, and thus the true radial profile and inner edge could be sharper than estimated below. Nevertheless, even with asymmetries we find sharp inner edges for those two systems that are inconsistent with collisional evolution (see below).

Having a better idea of the underlying intensity profiles, we identified the simplest parametric model that could reproduce the profiles derived by \textsc{frank} shown in Figure \ref{fig:FrankProfiles}. To decide which parametric model was most appropriate, we initially tested a series of different parametrizations and fitted them to the \textsc{frank} profiles. The final models are the ones that could reproduce the main significant features (local minima or maxima and inner and outer edge steepness) with the least number of free parameters, and these models are described in the following sub-sections from \ref{sec:q1erimodel} to \ref{sec:aumicmodel}.


All models have an inner section in the their surface brightness profiles that is parametrized as a power law  with an exponent $\alpha_{\rm i}$. Since debris discs are optically thin and their mm emission in the Rayleigh-Jeans regime, the surface density slope in this inner section is simply $\alpha_{\rm i}+1/2$ (a derivation for this can be found in Appendix \ref{app::inneredgederivation}). Therefore, whilst we fit the surface brightness inner slopes (found in table \ref{tab:allparameters}), all values referred to as the inner slope and discussed hereafter are $\gamma$ values (including those extracted from the literature). This inner section ends at a transition radius defined as the radial distance where the slope of the disc changes considerably (plateau's or starts decreasing). Beyond this transition radius, the disc follows a second power law (whether that is a middle or outer power law depends on the disc). As shown in Figure~\ref{fig:surface_density_model}, if the surface density profile is set by collisional evolution alone the transition from these two regimes should be smooth, and thus we introduce a smoothing exponent $\eta$, with higher values of $\eta$ making the transition abrupter (see description below). The smooth transition in the collisional model presented in Figure~\ref{fig:surface_density_model} is best fit with $\eta\approx2$. The parametric models were then Fourier transformed and sampled at the same $u-v$ points as the binned visibilities. Finally, the stellar flux was also included in our models as a free parameter---a point source at the origin becomes simply a real constant in the visibility space.

\begin{figure*}
    \includegraphics[width=0.49\textwidth]{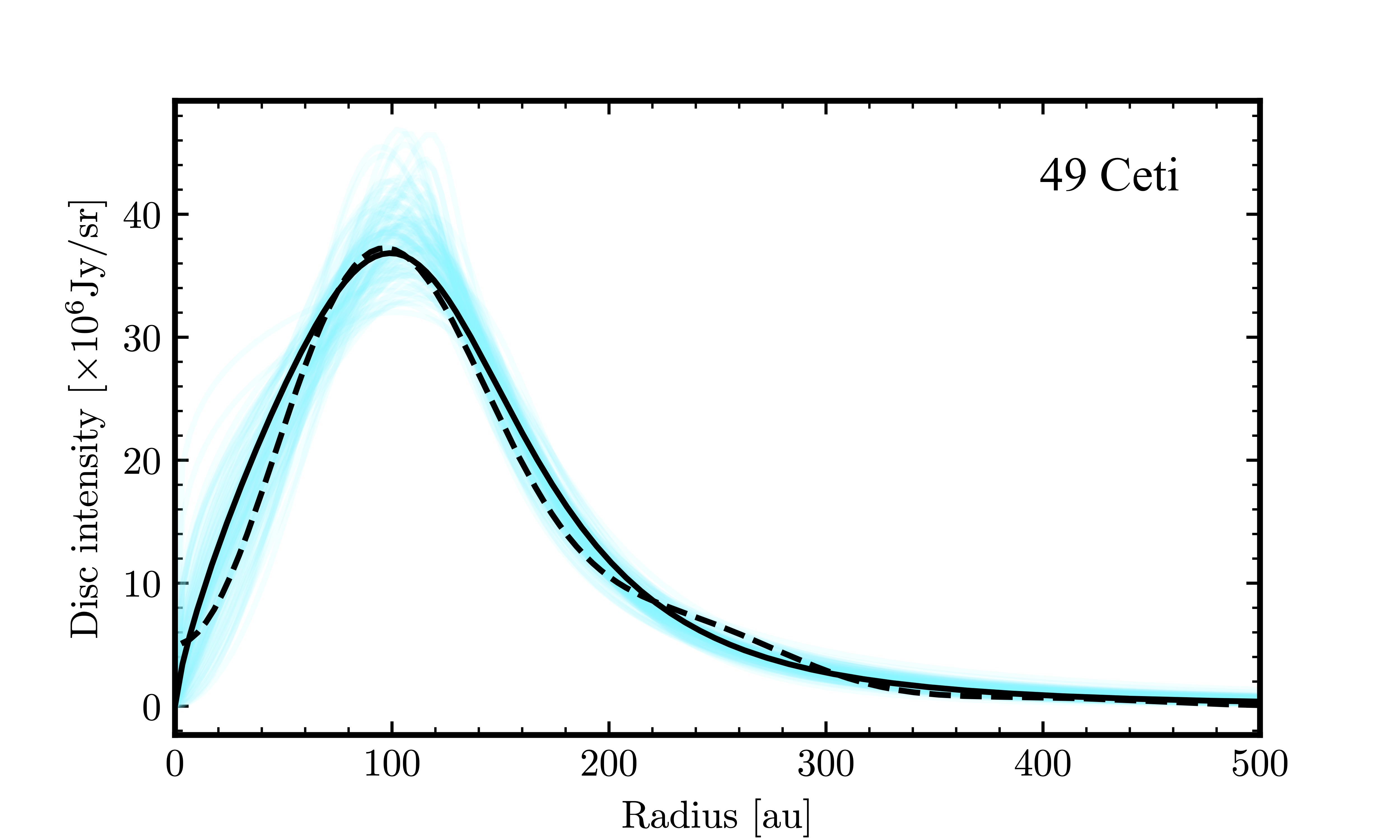}
    \includegraphics[width=0.49\textwidth]{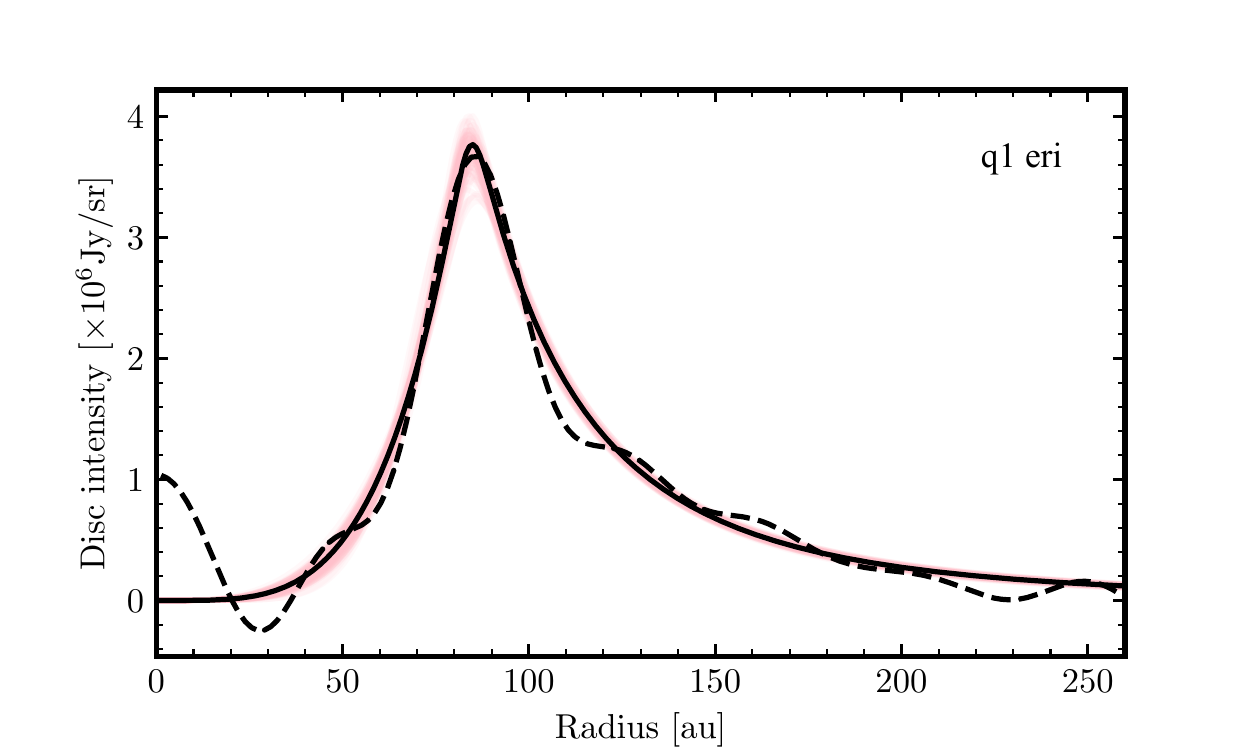}
    \includegraphics[width=0.49\textwidth]{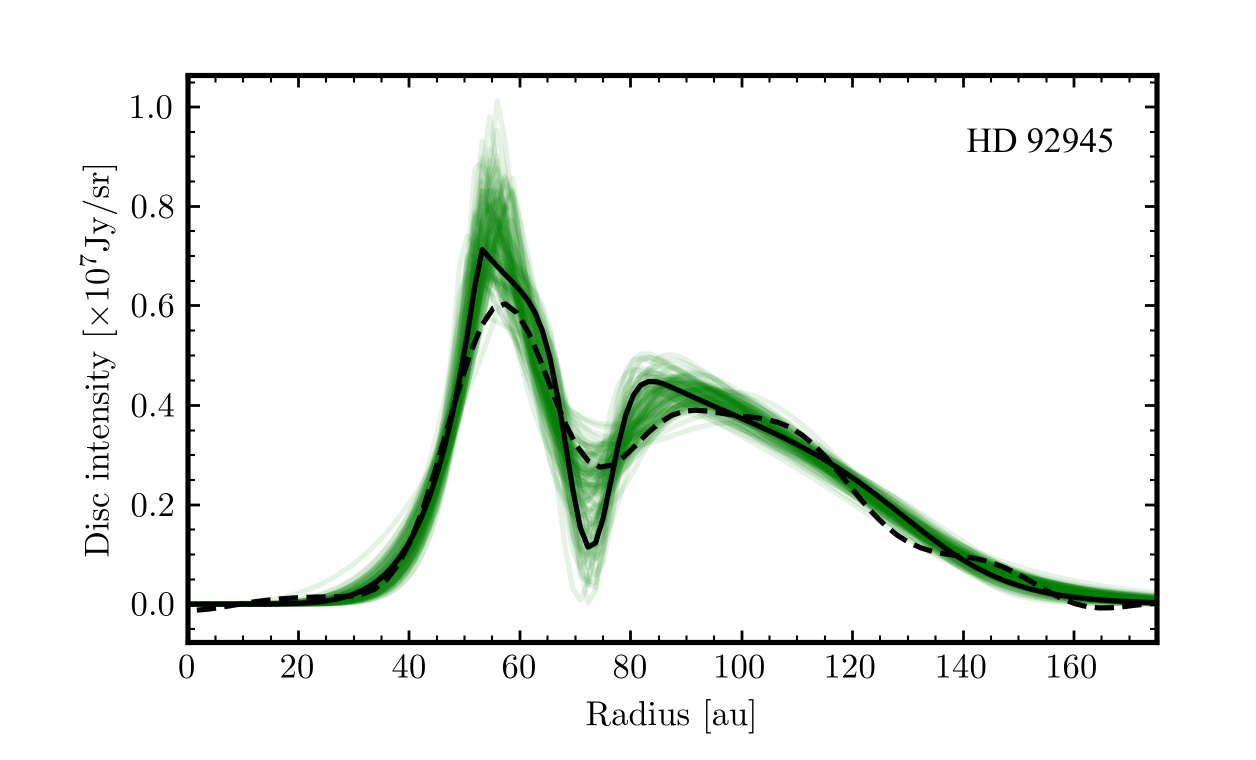}
    \includegraphics[width=0.49\textwidth]{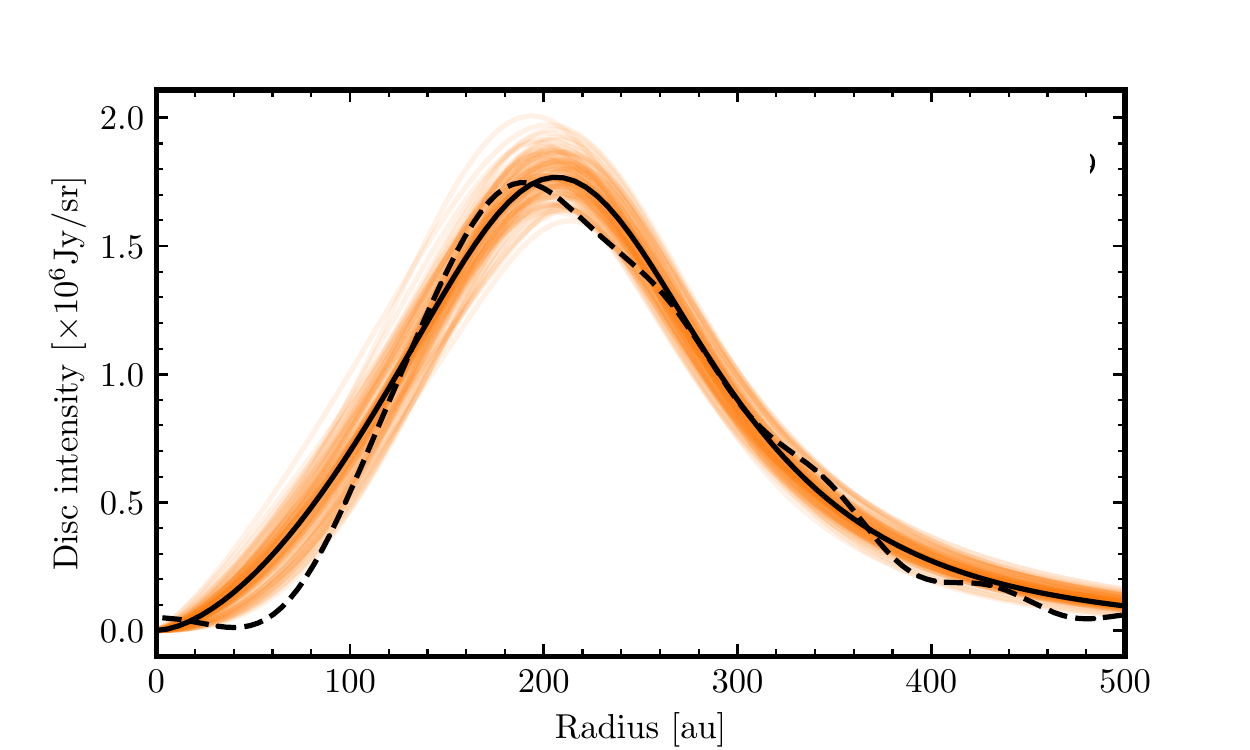}
    \includegraphics[width=0.49\textwidth]{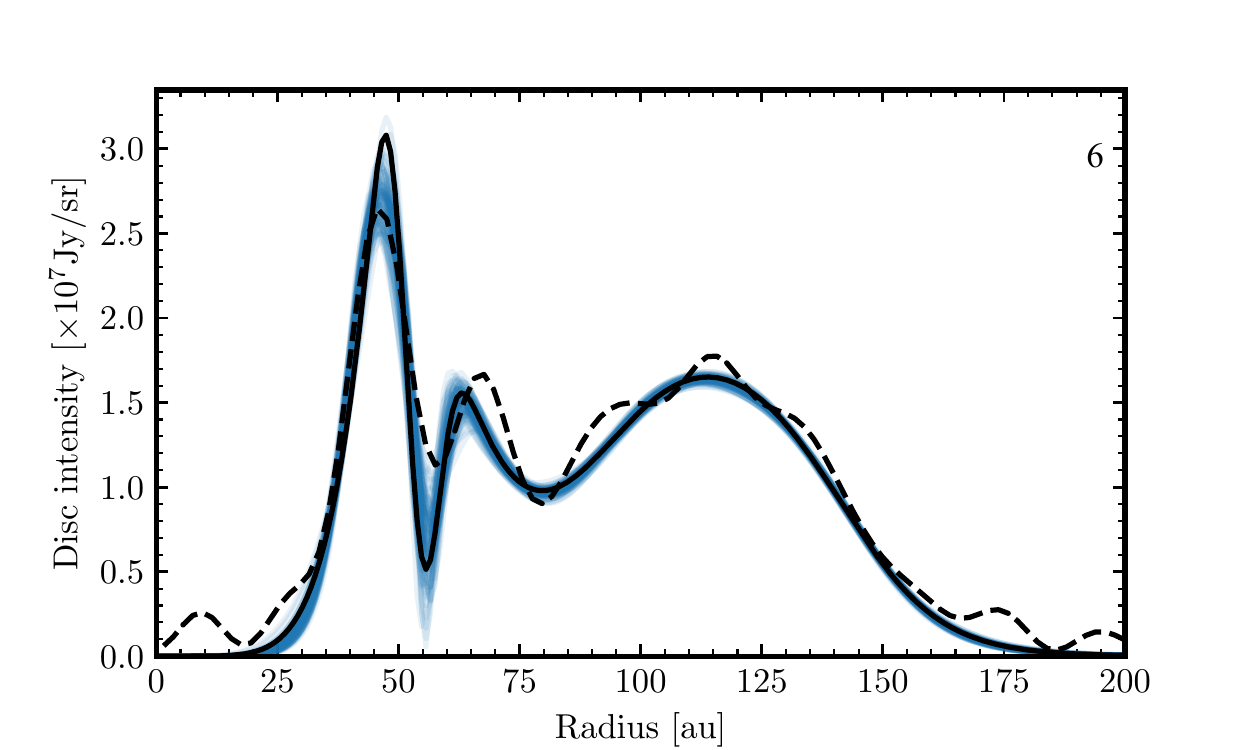}
    \includegraphics[width=0.49\textwidth]{HD206893VisibilitiesRadialProfile.pdf}
    \includegraphics[width=0.49\textwidth]{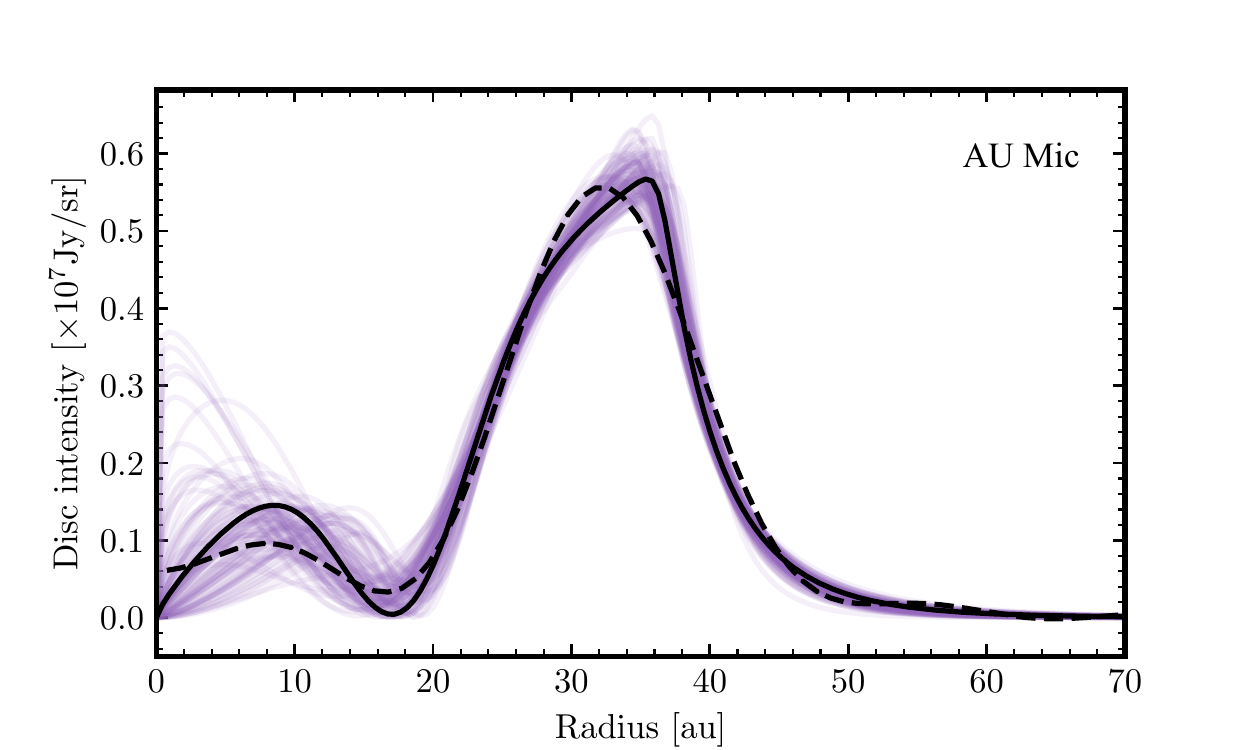}
\caption{Recovered surface brightness profiles from our parametric model fitting procedure. The solid black lines represent the best fit that minimizes the $\chi^2$. The dashed lines represent the \textsc{frank} profiles, for comparison with the fit found. The presented profiles correspond to band 6 (1.3~mm) for AU~Mic, band 7 (0.9 mm) for q$^1$~Eri, HD~92945, HR~8799, HD~107146, HD~206893, and band 8 (0.6~mm) for 49~Ceti. Note that for systems with where we use multiple bands, we fit both simultaneously allowing for a different $f_{\rm disc}$ and $f_{\star}$ for each band. The coloured lines represent the intensity profile of a random sample of 50 points from the posterior distribution of each system.}
\label{fig:ParametricProfiles}
\end{figure*}

\begin{figure}
    \centering
    \includegraphics[width=1\columnwidth]{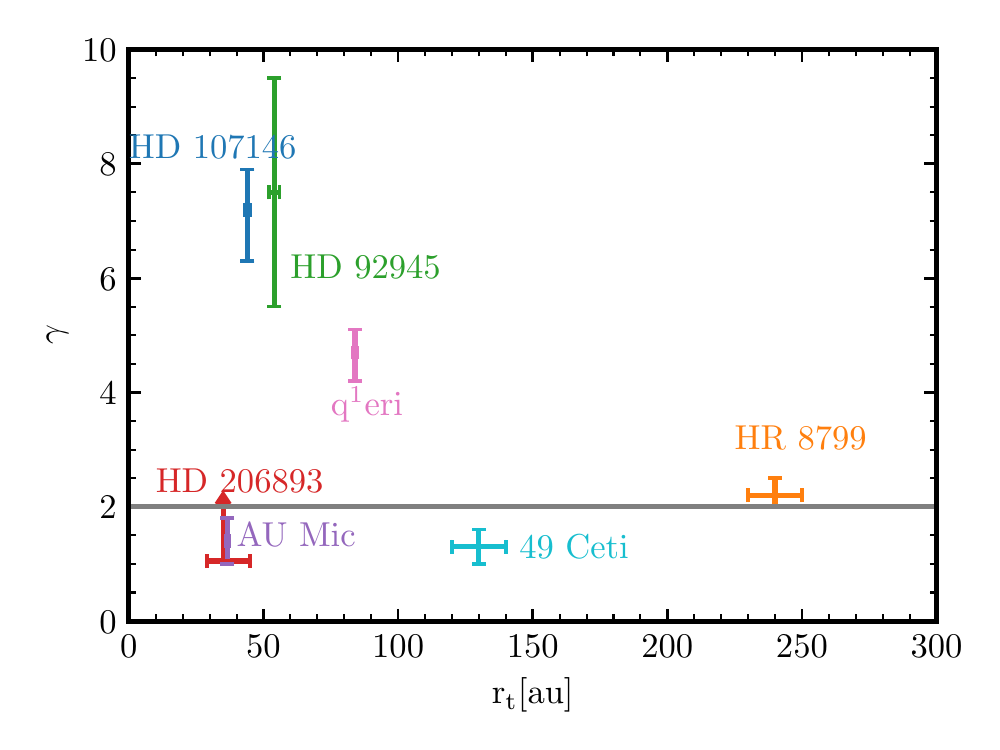}
    \caption{Values of the inner surface density slope ($\gamma$) against the location of the transition radius ($r_{\rm t})$ for the seven systems studied. The grey horizontal line shows the expected slope value of 2 if the inner slope is set by collisional evolution alone.}
    \label{fig:gammarin}
\end{figure}

The model visibilities were then compared directly to the binned visibilities by calculating the corresponding $\chi^2$. We find the best fit parameters and associated uncertainties by using the {\sc Python} package {\sc emcee} \citep{GoodmanWeare2010, Foreman-Mackey2013}, which implements an Affine Invariant MCMC Ensemble sampler to recover the posterior distribution of parameters. We assume uniform priors for each parameter and limited their range in a few cases to allow only physical solutions. We run the MCMC with 200 walkers and 2000 iterations, which we found was enough to ensure convergence (visually determined) and that the parameter space was well sampled. Figure~\ref{fig:ParametricProfiles} shows the recovered profiles of the 7 studied systems using our parametric models along the profile recovered by \textsc{frank}. Below we describe the model and results for each system. In Figure~\ref{fig:gammarin} and Table~\ref{tab:inneredges}, we summarise the values derived for the inner surface density slope as well as the estimated dust surface density and collisional lifetime of cm-sized grains. 

\begin{table*}
\centering
\caption{Inner surface density slope ($\gamma=\alpha_{\rm i}+1/2$), transition radius ($r_{t\rm}$), dust surface density at $r_{t\rm}$, estimated collisional lifetime of cm-sized dust using Equation~\ref{eq:fulltaucol}, and the ratio between their collisional lifetime and age of the system. HD~206893 the posterior distribution of $\gamma$ reached our upper bound of 10, and thus we report a $3\sigma$ lower limit instead.}
\label{tab:inneredges}
\begin{tabular}{lcccccc}
\hline
\hline
System      & Inner surface density slope $\gamma$   & Transition radius $r_{\rm t}$  [au]  & Dust surface density at $r_{\rm t}$ [$M_\oplus$~au$^{-2}$]  & $t_{\rm col}$(1~cm) [Myr] & $t_{\rm col}$(1~cm)$/t_{\rm age}$  \\ 
\hline
HR~8799      & $2.2^{+0.3}_{-0.2}$ & $240^{+10}_{-10}$ & $5.2\times10^{-7}$ & 9 & 0.2   \\
q$^1$~Eri   & $4.7^{+0.5}_{-0.4}$ & $84^{+1}_{-1} $ & $7.9\times10^{-7}$ & 0.8  & $6\times10^{-4}$    \\
HD~92945     & $7.5^{+1.7}_{-1.6}$        &  $54^{+2}_{-2}$ & $ 1.9\times10^{-6}$ & 0.2  &    $7\times10^{-4}$      \\
HD~107146 & $7.2^{+0.9}_{-0.7}$ &$44^{+1}_{-1}$  & $4.3\times10^{-6}$ & 0.04  &   $2-5\times10^{-4}$ \\ 
HD~206893 & $>1.05$ &$35^{+7}_{-10}$  & $9.5\times10^{-7}$ & 0.06 & $4\times10^{-4}$ \\ 
49~Ceti & $1.3^{+0.3}_{-0.3}$ & $130^{+10}_{-10}$  & $1.9\times10^{-6}$  & 0.4 & 0.01  \\ 
AU~Mic & $1.4^{+0.4}_{-0.4}$ & $36.5^{+0.7}_{-0.7}$ & $5.3\times10^{-6}$ & 0.05 & $2\times10^{-3}$ \\ 
\hline
\end{tabular}
\end{table*}

\subsubsection{q$^{1}$~Eri}
\label{sec:q1erimodel}
For this system \textsc{frank} found a profile with no visible gap in both band 6 and 7. To reproduce a similar morphology we decided to use a parametric model for the surface brightness $I(r)$ composed of two power laws joined at the disc peak or transition radius ($r_{\rm t}$)
\begin{equation}
     I(r)=I_0\left(\left(\frac{r}{r_{\rm t}}\right)^{-\eta\alpha_{\rm i}} + \left(\frac{r}{r_{\rm t}}\right)^{-\eta\alpha_{\rm o}}\right)^{-\frac{1}{\eta}},
    \label{eq:2plaw}
\end{equation}
where $\alpha_{\rm i}$ and $\alpha_{\rm o}$ are the power law exponents interior and exterior to $r_{\rm t}$,  $\eta$ determines how smooth or sharp the transition is, and $I_0$ a normalization factor. This was the best parametric model found for the visibilities and that could reproduce well the profile extracted by \textsc{frank}. In Figure \ref{fig:ParametricProfiles} we present the best-fit parametric model for q$^{1}$~Eri. We find a inner surface density slope $\gamma=4.7^{+0.5}_{-0.4}$. This is somewhat less steep, but still consistent within the errors to what was found by \citet{Lovell2021} that found a value of $>5.1$. The radial profiles from the parametric model and \textsc{frank} are consistent with each other, and the residuals are consistent with pure noise. In particular, the inner sections are very similar. The main differences are due to the non-significant wiggles caused by noise. We tested several parametric models and found that the derived values of the inner surface density slope were consistent across them. 

\subsubsection{HR~8799}
\label{sec:HR8799model}

Similar to q$^{1}$~Eri, the best parametric model for HR~8799 was found to be a model made of two power laws that join at the disc peak or transition radius as described by Equation~\ref{eq:2plaw}. The \textsc{frank} profile and the best fit in Figure \ref{fig:ParametricProfiles} coincide well and the residuals for this fit were also consistent with pure noise, thus this is a good fit for this disc. The inner surface density slope of HR~8799 was found to be $2.2^{+0.3}_{-0.2}$, which is consistent with the value found by \cite{Faramaz2021} of $3.0^{+0.9}_{-0.5}$. More importantly, this result confirms their findings that this disc does not have a well defined inner edge as expected if it was simply truncated by the HR~8799~b at its current location near 70~au \citep{Read2018}, but a surface density profile that smoothly rises with radius as expected in a collisional evolution scenario. Note that the uncertainties of our measured slope are a factor $\sim3$ smaller than the ones from \cite{Faramaz2021}. This difference is due to the model used by \citeauthor{Faramaz2021}, which consisted of a triple power-law. That model has many degeneracies that increase the uncertainty of the inner slope (see their Figure 9). Finally, the derived radial profile peaks at approximately 200~au rather than at $r_{\rm t}=240$~au. This difference is due to the low value of $\nu$ (<1.6) making the profile smoother and the outer slope that is steeper than the inner one (1.7 vs -4.4).

\subsubsection{49~Ceti}

For 49~Ceti the best parametric model was also found to be two power laws joining at the disc peak or transition radius, in the same manner as for q$^1$~Eri and HR~8799 and as described by Equation~\ref{eq:2plaw}. The \textsc{frank} profile and the best fit in Figure \ref{fig:ParametricProfiles} coincide well and the residuals for this fit were also consistent with pure noise, thus this is an appropriate model for this disc. The value for the inner surface density slope of 49~Ceti was found to be $1.3^{+0.3}_{-0.3}$ through this fitting. This is a very low value and consistent with collisional evolution and the fact that no massive planets have been found around 49~Ceti that could truncate or stir the disc. \citet[][]{Hughes2017} found an inner slope of $2.5^{+0.8}_{-2.2}$, which is consistent with our finding and with collisional evolution. Similar to HR~8799, the disk peaks at a radius slightly smaller than $r_{\rm t}$ due to the low value of $\nu$ and the outer slope being steeper than the inner one.

\subsubsection{HD~92945}
\label{sec:HD92945model}

For this system, \textsc{frank} found a wide disc with a single gap, and thus we chose a parametric model that could mimic this gap and adjust to the inner and outer edge sharpness. This model consists of a power law inner section, a middle power law, a Gaussian gap, and an outer edge parameterised as an hyperbolic tangent following \cite{Marino2021}
\begin{eqnarray}
     I(r)= I_0 G(r) \left(\left(\frac{ r}{ r_{\rm t}}\right)^{-\eta\alpha_{\rm i}}+\left(\frac{r}{r_{\rm t}}\right)^{-\eta\alpha_{\rm m}}\right)^{-\frac{1}{\eta}}\left(1+\tanh\left(\frac{r_{\rm out}-r}{l_{\rm out}}\right)\right),\label{eq:1gap} \\
    G(r) = 1-\delta_{\rm g} \exp\left({\frac{(r-r_{\rm g})^2}{2\sigma_{\rm g}^2}}\right).
\end{eqnarray}
where $\alpha_m$ is the slope of the middle section of the disc (if there was no gap), $r_{\rm out}$ is the location of the outer edge, $l_{\rm out}$ determines how smooth or sharp the outer edge is, and $G(r)$ represents a Gaussian gap centred at $r_{\rm g}$, with a standard deviation $\sigma_{\rm g}$ and a fractional depth $\delta_{\rm g}$. In Figure \ref{fig:ParametricProfiles} we present the best-fit parametric model for HD~92945. Again, the residuals were consistent with pure noise, which means that the chosen model is enough to explain the main features present in the data. The inner surface density slope was constrained to $7.5^{+2}_{-2}$, which is consistent with the lower limit of 5.7 derived by \cite{Marino2019}.

\subsubsection{HD~206893}

For this system, \textsc{frank} found a wide disc with a single gap, and thus we chose the same parametric model as for HD~92945 (described by Equation~\ref{eq:1gap}). In Figure \ref{fig:ParametricProfiles} we present the best-fit parametric model for HD~206893. Again, the residuals were consistent with pure noise, which means that the chosen model is enough to explain the main features present in the data. The recovered profile is very uncertain around the inner section, with it only managing to recover a lower limit for the slope. The best-fit model is in good agreement with the \textsc{frank} profile. Furthermore, it recovered the gap in the disc. The value of the inner slope found for HD~206893 is $>1.05$. Due to this being only a lower limit, the inner  slope is both consistent with being shallow (i.e. consistent with collisional evolution) and sharp (consistent with being truncated by planets), something that \citet{Marino2021} also found. This system is known to host two massive companions interior to the disc at semi-major axes of 3.5 and 9.7~au \citep{Milli2017a, Delorme2017, Hinkley2022}. The disc inner edge or transition radius and slopes are, however, very uncertain and thus it is hard to assess if the outer companion is what set the inner extent of the disc. 

\subsubsection{AU~Mic}
\label{sec:aumicmodel}
For AU~Mic, the best parametric model was found to be two power laws joining at the disc peak or transition radius and an additional Gaussian gap following
\begin{eqnarray}
    I(r)=I_0 G(r) \left(\left(\frac{r}{r_{\rm t}}\right)^{-\eta\alpha_{\rm i}} + \left(\frac{r}{r_{\rm t}}\right)^{-\eta\alpha_{\rm o}}\right)^{-\frac{1}{\eta}},
    \label{eq:2plawgap} \\
    G(r) = 1-\delta_{\rm g} \exp\left({\frac{(r-r_{\rm g})^2}{2\sigma_{\rm g}^2}}\right).
\end{eqnarray}
The \textsc{frank} profile and the best fit in Figure \ref{fig:ParametricProfiles} coincide well and the residuals for this fit were also consistent with pure noise, thus this is an appropriate fit for this disc. At a radius smaller than 10~au the shape is very uncertain (which is in agreement with what \textsc{frank} finds), however, all tested models required a local minimum around 20~au and significant emission at 10~au, thus making a gap in the disc a likely feature. The single gap model was compared to the no gap model, and considering the added number of parameters of the single gap model, the \textit{Bayesian Information Criterium} \citep[BIC,][]{Schwarz1978} value difference between the models is still $>10$ and thus statistically significant \footnote{The difference in the BIC value of the two models was 15, with the model with the gap having the lower BIC, supporting the usage of this model.}. The inner surface density slope of AU~Mic was found to be $0.9^{+0.4}_{-0.4}$. This result implies that the inner section is shallow and consistent with collisional evolution. Previous analyses by \citet[][]{Daley2019,Marino2021,Vizgan2022} also found a surface density profile that gently rises with radius, with \citeauthor{Daley2019} finding an inner slope of $0.9^{+0.5}_{-0.4}$ which is in good agreement with our findings.

\subsubsection{HD~107146}
\label{sec:HD107146model}

For HD~107146 we tested out various different models due to its complexity as \textsc{frank} revealed 2 gaps in the disc as opposed to the single shallow and wide gap found in previous analysis. The chosen parametric model for HD~107146 after extensive testing was a power law inner section, a middle section power law, a hyperbolic tangent outer edge, and two Gaussian gaps
\begin{eqnarray}
\nonumber     I(r)= I_0 G_1(r)G_2(r) \left(\left(\frac{ r}{ r_{\rm t}}\right)^{-\eta\alpha_{\rm i}}+\left(\frac{r}{r_{\rm t}}\right)^{-\eta\alpha_{\rm m}}\right)^{-\frac{1}{\eta}}\\ 
     \left(1+\tanh\left(\frac{r_{\rm out}-r}{l_{\rm out}}\right)\right),\label{eq:2gap} \\ 
    G_1(r) = 1-\delta_{\rm g1} \exp\left({\frac{(r-r_{\rm g1})^2}{2\sigma_{\rm g1}^2}}\right),\\
    G_2(r) = 1-\delta_{\rm g2} \exp\left({\frac{(r-r_{\rm g2})^2}{2\sigma_{\rm g2}^2}}\right),
\end{eqnarray}
This was decided by comparing the BIC values for the single gap model against the double gap, and when considering the increased complexity of the two gap model the BIC value was still better for this model \footnote{The difference in their BIC values was 37, with the two gap model having a lower BIC, and thus supporting the use of this model.}. The fit can be seen in Figure \ref{fig:ParametricProfiles}. The fit is slightly different to the \textsc{frank} profile, as the first gap is deeper and narrower  than in the \textsc{frank} profile. However, various models were tested and they all had a consistent inner section. Furthermore, whilst the shapes of the two gaps in the disc are uncertain, the fact that both \textsc{frank} and the parametric fit both converge to a two gap disc strongly supports the presence of substructures within the broad gap. Moreover, the double gap is also recovered from Clean images from the newest and highest resolution data, although at a lower significance partly due to the lower resolution of Clean images compared to deconvolved models (Appendix~\ref{app:data}). The inner surface density slope was found to be $7.2^{+0.9}_{-0.7}$. \cite{Marino2018b} finds an inner slope of $11.6^{+3.0}_{-2.7}$, which appears to be much higher than the value found in this investigation, however still consistent within $3\sigma$. Both derived values are much higher than the value of 2 expected in a pure collisional evolution scenario, which is what we aimed to determine. Regarding the two gaps, we found that these are centred at $56\pm1$~au and $79\pm1$~au; these results will be examined in more detail in a future work using JWST data from cycle 1.


\section{Discussion}
\label{sec:discussion}
\subsection{Surface density}
\label{dis:surface_density}
Based on the derived values for the inner surface density slope $\gamma$ and the transition radius $r_{\rm t}$ presented in Table~\ref{tab:inneredges}, we can now assess if these could be consistent with a disc evolving through collisions without the need for it being truncated. We expect that the surface density slopes in the inner regions with unimpeded collisional evolution to be equal to $\gamma=0.75 q_g +0.75\approx2$ ($q_g=1.69$, \S\ref{sec:radialslope}). This value of 2 is consistent with the ones derived for HR~8799, HD~206893, 49~Ceti and AU~Mic. For the rest, we can rule out a smooth inner section as it would be produced by pure collisional evolution. In the following subsection we will use this information to constrain $x_{\rm MMSN}$ and $D_{\rm max}$ in these systems. A caveat to keep in mind when comparing the values of $\gamma$ derived from observations with our model, is that it assumes that the timescale it took solids to be stirred and initiate the collisional cascade (at all radii) is much shorter than the collisional timescale of the larger bodies and the age of the system. If the stirring timescale was longer than the collisional timescale, pure collisional models could produce a sharp inner edge \citep{Kennedy2010}.

Based on the  surface brightness at the transition radius, we can estimate the collisional lifetime of the observed mm-sized grains and compare this to the age of these systems to assess whether the dust is being replenished by collisions of larger solids. To do this, we convert the dust surface brightness into a dust surface density assuming a dust temperature equal to the equilibrium temperature at the transition radius and a dust opacity $\kappa_{\rm D<1\ cm}=1.6 (\lambda/1\ \mathrm{mm})^{-0.9}$ cm$^{2}$~g$^{-1}$ calculated using Mie Theory for a grain size distribution up to 1~cm \citep[][and references therein]{Marino2018b}. We then use Equation~\ref{eq:fulltaucol} to compute the lifetime of 1 cm-sized grains (replacing $q_{\rm g}$ by $q_{\rm s}$ and setting $\epsilon=1$ to be valid for $D_{\max}=1$~cm $<D_{\rm b}$). Table~\ref{tab:inneredges} presents the estimated dust surface densities (4th column) and the collisional lifetime of cm-sized grains in Myr (5th column) and relative to the age of the systems (6th column). We find that the lifetime of grains at the transition radius is much shorter than their ages for all the discs, except for HR~8799 (9~Myr) where their lifetime is shorter but still comparable to the age of the system (especially considering the multiple uncertainties when transforming the disc surface brightness into a density). This indicates that, apart from HR~8799, the dust in these systems is collisionally processed and its replenishment requires the presence of larger solids.

\subsection{Constraints on $x_{\rm MMSN}$ and $D_{\rm max}$}

\begin{figure}
\centering
\includegraphics[width=1.0\columnwidth]{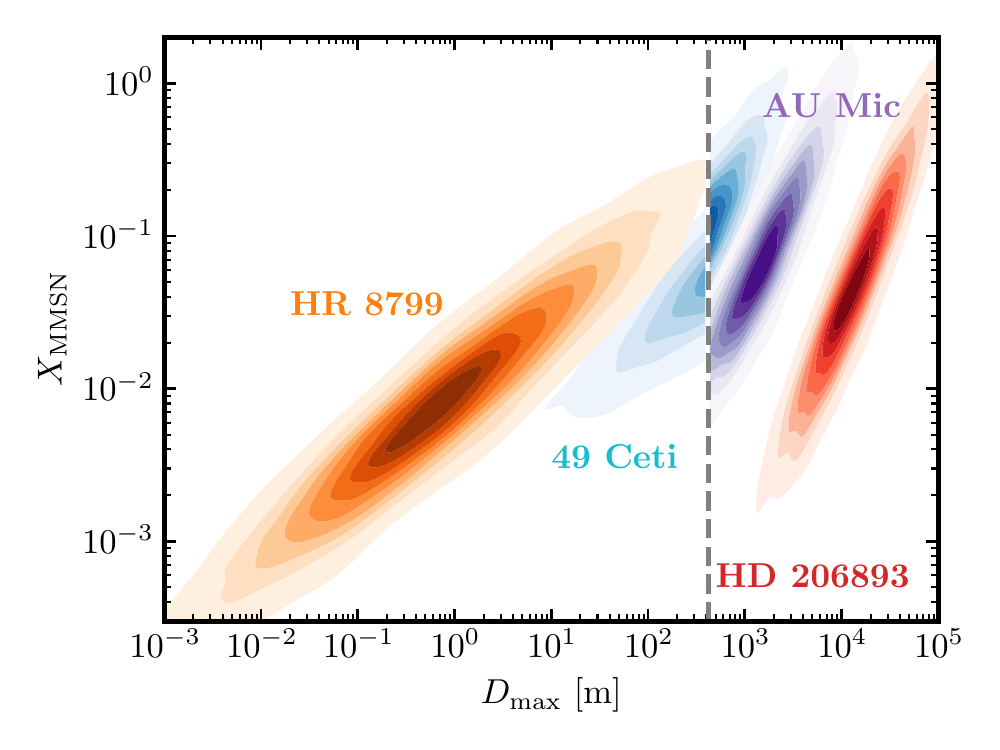}
    \caption{Constraints on $x_{\rm MMSN}$ and $D_{\rm max}$ for four discs with inner edge slopes consistent with collisional evolution. The filled contours show the most likely values when considering the systematic uncertainties in the system parameters. The faintest contours represent the 95\% confidence limit. The vertical grey dashed line represents $D_{\rm b}$ where there is a discontinuity in the model.}
    \label{fig:xMMSNvsDmax_smooth}
\end{figure}

Based on the inner slopes and transition radii derived above, we now proceed to use this information to constrain $x_{\rm MMSN}$ and $D_{\rm max}$. We start by focusing on HR~8799, HD~206893, 49~Ceti and AU~Mic, which have inner sections that are consistent with being shallow and shaped by collisional evolution as discussed above. Using equations \ref{eq:dmaxgbetterunits} and \ref{eq:xmmsngbetterunits} (or \ref{eq:dmaxsbetterunits} and \ref{eq:xmmsnsbetterunits} if $D_{\max}<D_{\rm b}$), the systems' parameters in Table~\ref{tab:basicdatasystems}, the transition radius as a proxy for the critical radius $r_{\rm c}$ (found in Table \ref{tab:inneredges}), and the estimated dust surface density, we derive $x_{\rm MMSN}$ and $D_{\rm max}$ for these four systems. In order to account for the different systematic uncertainties in the system parameters and emitting properties of dust, we perform a Monte Carlo simulation injecting noise with a log-Normal distribution to the dust surface brightness (0.3 dex), $r_{\rm c}$ (0.1 dex), $t_{\rm age}$ (0.2 dex), $M_{\star}$ (0.05 dex). The resulting distributions of $10^{4}$ points are presented in Figure~\ref{fig:xMMSNvsDmax_smooth}, with the faintest filled contours representing the 95\% confidence limit. Note that there is a discontinuity at $D_{\max}=D_{\rm b}$, where none of the equations used are strictly valid. 

Overall we find that the values for the transition radii and dust surface densities can be explained by sub-km planetesimals for three systems. AU~Mic, 49~Ceti and HR~8799 require $D_{\max}$ in the range $3\times10^{2}-10^{4}$, $10-10^{3}$ and $10^{-3}-10^{2}$~m, respectively. These values for $D_{\max}$ are much lower than the typical values assumed for debris discs (10-1000~km). The large size and short age of these systems means that they do not require large planetesimals to sustain their dust levels. HD~206893 with a likely older age compared to the rest requires larger planetesimals with a size between 1-100~km. If we look at $x_{\rm MMSN}$ we find that the three systems require values below 1 (i.e. surface densities of solids that are lower than the MMSN). This means that the total mass in solids is not in an obvious contradiction with the available solid mass in protoplanetary discs, avoiding the disc mass problem \citep{Krivov2018, Krivov2021}.

It is interesting to note that HR~8799 does not require large solids to sustain the observed dust levels. In other words, the circumstellar dust is consistent with being simply a leftover from the protoplanetary disc phase. This conclusion is consistent with the long lifetime of cm-sized grains that we estimated in \S\ref{dis:surface_density}. The derived dust mass below 1~cm is just $\sim0.1\ M_{\oplus}$, which would have been $\lesssim0.1\%$ of the dust mass present in its primordial protoplanetary disc (assuming a disc mass of 0.05~$M_{\odot}$ and a gas-to-dust ratio of 100). This leftover dust could have been the small fraction that did not grow to pebble sizes fast enough to radially drift towards the star or a pressure maximum near HR~8799 b's orbit.

These conclusions on $x_{\rm MMSN}$ and $D_{\rm max}$, however, rely on the assumption that the disc inner edge was shaped by collisional evolution and not by other processes such as planet-disc interaction. Such interactions could be the ones responsible for shaping HR~8799 and HD~206893 inner edges \citep{Faramaz2021, Marino2021}. Therefore, these results are only valid under a pure collisional evolution scenario. Note that as discussed in \S\ref{sec:rcequations} our analytic model is likely over-predicting the dust levels by a factor $\sim3$. If we take this into account, the required $D_{\max}$ values to explain observations would be a factor $\sim3$ larger for $D_{\max}>D_{\rm b}$ and a factor $\sim30$ larger $D_{\max}$ for $D_{\max}<D_{\rm b}$. Similarly, $x_{\rm MMSN}$ would be a factor $\sim10$ larger in both regimes. Therefore, the derived values must be taken with caution.

We can now focus on the opposite scenario: the size distribution is not yet in collisional equilibrium throughout the disc as the largest bodies have not collided yet. In this scenario the inner sections could be much steeper than our collisional evolution model predicts (thus has a $\gamma>2.0$ as HD~107146, HD~92945, and q$^{1}$~Eri) or be smooth due to something other than collisional evolution \citep[e.g. very high eccentricities,][]{Marino2021}. There are two conditions that we can use to constrain $x_{\rm MMSN}$ and $D_{\rm max}$ assuming this scenario is true. First, $r_{\rm c}$ must be smaller than $r_{\rm t}$. Otherwise, we would see a slowly increasing surface density from $r_{\rm t}$ to $r_{\rm c}$. This condition can be implemented using Eq.~\ref{eq:rcbetterunits} to derive the maximum $x_{\rm MMSN}$ as a function of $D_{\rm max}$ such that $r_{\rm c}<r_{\rm t}$. The second condition is that the combination of $x_{\rm MMSN}$ and $D_{\rm max}$ must reproduce the estimated dust surface density. To implement this second condition we use Eq.~\ref{eq:sigmadustbetterunits} to constrain $x_{\rm MMSN}$ as a function of $D_{\rm max}$ such that it matches the observed dust surface density. This assumes $\alpha=3/2$ and that the largest body in collisional equilibrium is larger than $D_{\rm b}$ \citep{Marino2017b}. This is not valid for HR~8799 and 49~Ceti given their large size and young age, which are consistent with $D_{\max}<D_{\rm b}$\footnote{This was confirmed using numerical simulations from \cite{Marino2017b}.}. Therefore, these two systems are excluded from this analysis. 

\begin{figure}
\centering
\includegraphics[width=1.0\columnwidth]{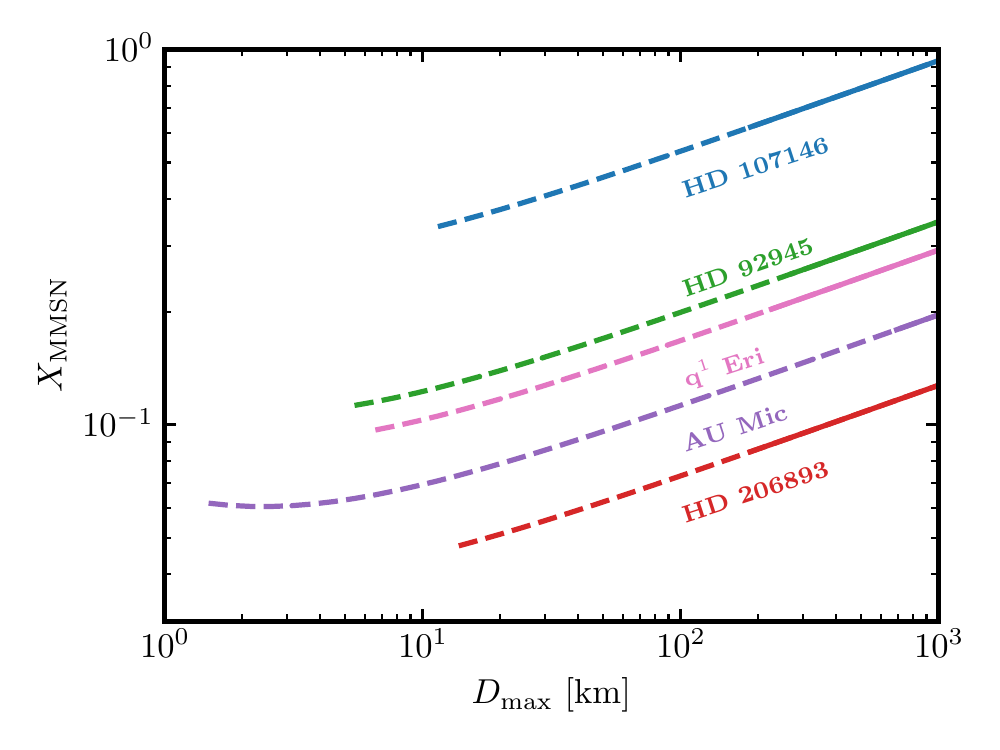}
    \caption{Constraints on $x_{\rm MMSN}$ and $D_{\rm max}$ for 5 discs for which  $D_{\max}\geq D_{c}>D_{\rm b}$, 3 of which have inner slopes inconsistent with collisional evolution (HD~107146, HD~92945 and q$^{1}$~Eri). The coloured lines show the required $x_{\rm MMSN}$ as a function of $D_{\rm max}$. The solid lines mark the range where the surface density and size of the largest planetesimals would be enough to stir the disc (self-stirring). Note that the lines are restricted to the range in which $x_{\rm MMSN}$ is below the maximum value such that $r_{\rm c}<r_{\rm t}$.} 
    \label{fig:xMMSNvsDmax_sharp}
\end{figure}

Figure~\ref{fig:xMMSNvsDmax_sharp} shows the required $x_{\rm MMSN}$ to explain the observed amount of dust as a function of $D_{\rm max}$ (second condition). We only plot $D_{\rm max}$ for which $r_{\rm c}<r_{\rm t}$ (first condition). All systems are consistent with $x_{\rm MMSN}\lesssim1$ (avoiding the disc mass problem), and all except HR~8799 and 49~Ceti require $D_{\max}>D_{\rm b}$ (i.e. in the gravity regime). We can add as a third condition that the surface density and maximum planetesimal sizes are large enough to have self-stirred the disc within the age of the system \citep{Krivov2018selfstirring}. This third condition is met along the solid section of the lines in Figure~\ref{fig:xMMSNvsDmax_sharp}. We find that the discs could be self-stirred in HD~107146, HD~92945, HD~206893, and q$^1$~Eri if $D_{\max}\gtrsim 100$~km. Note, however, that the equations used to derive these lines might be invalid for $D_{\max}>100$~km since $X_{\rm c}$ could be larger than 1 (\S\ref{sec:collrate}). HD~107146 stands out in this figure for requiring the largest value of $x_{\rm MMSN}$ between $0.3-1$, which given its radial span from 44 to 144~au, is equivalent to a total mass $\sim20-60\ M_{\oplus}$. As noted before, our model over-predicts the dust surface density level, and thus the required $x_{\rm MMSN}$ values could be a factor $\sim10$ larger. Even with this correction, we find that all these discs can avoid the disc mass problem. 


Note that when deriving $x_{\rm MMSN}$ and $D_{\max}$ using the equations in \S\ref{sec:rcequations} we have assumed a particular solid strength law $Q^{\star}_D$ corresponding to ice \citep{BenzAsphaug1999} as the solids at tens of au in these systems would probably be similar to Solar System comets in composition. Ice has one of the weakest strengths, and thus if we had assumed stronger solids the derived values of $D_{\max}$ and $x_{\rm MMSN}$ would be lower. For example, assuming the strength values of basalt \citep{BenzAsphaug1999} that yield a similar $D_{\rm b}$ and a $Q^{\star}_{D_{\rm b}}$ value a factor 2 higher, we find that the $D_{\max}$ and $x_{\rm MMSN}$ values derived from Figure~\ref{fig:xMMSNvsDmax_smooth}  are a factor $400$ and 5 smaller, respectively, in the strength dominated regime ($D_{\max}<500$~m). In the gravity dominated regime ($D_{\max}>500$~m), we find $D_{\max}$ and $x_{\rm MMSN}$ values a factor $10$ and 5 smaller, respectively. In addition, the values of $x_{\rm MMSN}$ derived for Figure~\ref{fig:xMMSNvsDmax_sharp} would be a factor 3 smaller. Therefore, stronger solids would imply an even smaller maximum size in the collisional cascade and lower solid surface densities and thus it does not alter our general conclusions of small planetesimals.

\subsection{Truncation by planets}

Here we investigate the masses and locations of putative planets that could have truncated the discs in order to result in the shapes found. Following \cite{Pearce2022}, we find the minimum planet mass using their publicly available code\footnote{\url{https://github.com/TimDPearce/SculptingPlanet}}, which accounts for the inner edge location and the scattering time being shorter than the age of the system. We define the inner edge location as the radius at which the intensity (recovered by our parametric model fits) reaches half of the value at the top of the first peak in the radial profile. Note that \cite{Pearce2022} had estimated these masses using slightly different inner edge values collected from the literature at that time. Therefore, here and in the following section we repeat this exercise using our derived inner edge locations. We assume an eccentricity of zero as the discs are all approximately axisymmetric, and thus set the apocentre and pericentre of the discs inner edge to be equal. Using this code and these assumptions, we find the minimum planet masses and maximum semi-major axis of the planets found in Table \ref{tab:massresults}. 

\begin{table*}
\centering
\caption{Values for the minimum planet mass for single planet truncation to occur ($M_{\rm p}$), maximum semi-major axis of the planet for truncation to occur ($a_{\rm p}$), minimum multi-planet mass for truncation to occur ($M_{\rm p,n}$), and minimum planet mass for stirring to occur ($M_{\rm p, stir}$) for all of the systems. All of the minimum masses are calculated using the inner and outer edges derived from our modelling, where the inner edge is the radius at which the intensity (recovered by our parametric model fits) gets to half of the value at top of the first peak and the outer edge is the radius at which the intensity is half of the value of the outermost peak.}
\label{tab:massresults}
\begin{tabular}{lcccccc}
\hline
\hline
System   & Inner Edge Value [au] & Outer Edge Value [au] & $M_{\rm p} [M_{\rm Jup}]$ & $a_{\rm p} [\rm au]$ & $M_{\rm p,n} [M_{\oplus}]$ & $M_{p,\rm stir} [M_{\oplus}]$ \\ 
\hline
HR~8799  &109 & 300 & $1.7$       &     $80$     & 65         &    650       \\
q$^1$eri  & 70 & 106 & $0.20$       &  $59$        &    0.88           &    0.60          \\
HD~92945  & 48 & 124& $0.26$      &   $39$        & 2.1      &    16         \\
HD~107146 & 40 & 143 &  $0.39$       &   $32$       & 3.8      &    120         \\ 
HD~206893 & 31 & 138 & $0.34$        &   $26$      &    2.3     &        200    \\ 
49~Ceti &   35 & 175&     $1.0$   &  $28$    &    15     &       2200      \\ 
AU~Mic &   25&39 &  $0.36$     &  $19$   &    8.0    &  6.3     \\ 
\hline
\end{tabular}

\end{table*}

The estimated minimum planet masses for each of the systems can be found in Table \ref{tab:massresults}. The masses of the single planet truncation found ranged from $0.2\ {\rm M_{\rm Jup}}$ for q$^1$~Eri to $1.7 {\rm M_{\rm Jup}}$ for HR~8799. Most of these values are well below the existing constraints for these systems and beyond the current capabilities of ground-based direct imaging instruments which can only detect planets more massive than few Jupiter masses \citep[e.g.][]{Nielsen2019, Langlois2021}. Only for HR~8799 and 49~Ceti, the minimum planet masses are very close to the detection limits. For 49~Ceti, SPHERE observations could have detected a 2~$M_{\rm Jup}$ planet near 95~au \citep{Choquet2017}, but not if that planet was near the minor axis of the disc and at a much smaller apparent separation. For HR~8799, SPHERE observations have ruled out the presence of a 0.6~$M_{\rm Jup}$ or more massive planet beyond 100~au \citep{Zurlo2022} and thus a fifth planet responsible for truncating the the disc by itself would have been detected. However, it has been suggested that the known four planets in the HR~8799 system (the outermost at 70~au) migrated inwards into its close to resonant configuration and thus could have truncated the disc without the need of an additional planet \citep[e.g.][]{Gozdziewski2018}. 

It is also possible that the inner edge is truncated by a multi-planet system \citep[][]{Shannon2016}. Using Equation 15 from  \cite{Pearce2022}, we calculate the minimum mass of planets required to clear their orbits in a multi-planet system within the age of the system assuming the outermost planet is at the disc inner edge. We have calculated these masses for all of the systems, even those that have been found to have a shallow inner edge consistent with collisional evolution, since even in those systems planets may have been responsible for truncating the disc. The range of multi-planet masses found is between $1 {\rm M_{\oplus}}$ and $65 {\rm M_{\oplus}}$, all below the range of detection for ground-based direct imaging instruments. JWST will allow for detection of planets $\gtrsim0.1 {\rm M_{\rm Jup}}$ \citep{Carter2021}, therefore some of these putative planets could be directly imaged in the near future. In fact, the 7 systems studied here will be directly imaged during JWST's cycle 1. Finally, 4 of these systems (HR~8799, HD~92945, HD~107146 and HD~206893) have significant Gaia eDR3proper motion anomalies \citep{Kervella2022}. For the case of HR~8799 and HD~206893 this is caused by one of the known planets in these systems \citep{Brandt2021, Hinkley2022}, whereas for HD~92945 and HD~107146 such companion has not been detected yet.


\subsection{Stirring by planets}
\label{sec:stirring}
Given the location of the disc inner edge and the extent of these discs, we can also estimate the minimum mass of a planet just interior to the disc inner edge for it to stir the orbits of solids across the whole extent of the disc. For this, we use Equation 23 in \citep{Pearce2022}, assuming a planet eccentricity of 0.1.
This is a necessary assumption to make as stirring requires $e_{\rm p}>0$, but the systems are approximately axisymmetric and thus we use a low $e_{\rm p}$. We define the outer edge radius as the location at which the disc intensity gets to half of the outermost peak. The results for the values of $M_{\rm p,stir}$ can be found in Table \ref{tab:massresults}. The minimum planet mass for stirring range from $0.6\ {\rm M_{\oplus}}$ for q$^1$eri to $2200\ {\rm M_{\oplus}}$ for 49~Ceti. The masses for HR~8799, HD~107146, HD~206893, and 49~Ceti are above the expected JWST detection limit, and these four systems have observing time allocated during cycle 1. This would serve to test models of planet stirring and disc truncation by planets.

\subsection{Limitations}

In addition to the limitations of our model described throughout the paper due to the approximations that we used, here we briefly describe a few additional caveats. First, our model assumes that the dynamic excitation or relative velocities of solids is not a function of grain size. This might not be true across the size distribution if collisional damping is important \citep[e.g.][]{PanSchlichting2012} and thus could slightly affect the surface density slope. Our assumption would also not be valid near the bottom of the collisional cascade where radiation pressure (or the effect of stellar winds) is not negligible. Near the blow-out size, the dynamical excitation will be set by radiation pressure (or stellar winds) and the smallest grains will be released onto very eccentric orbits. The effects of radiation pressure were considered by \cite{Schuppler2016} and they found a surface density of solids in the inner regions rising as $r^2$ (see their Figure 1). Note that the optical depth (shown in the same figure) has a different radial profile that rises more slowly, but this is dominated by the smallest grains that are affected by radiation pressure. Since in this paper we focus on the mm-sized grains we conclude that neglecting the effect of radiation pressure should not affect our conclusions.

A second limitation in our models is that we do not consider the effects of radial mixing when eccentricities are high or any radial transport. Our model treats the collisional evolution at each radii independently, which would not be valid for high orbital eccentricities as those orbits would span a wide range of orbital radii. Our model also neglects the effect of P-R drag (the main radial transport mechanism in the absence of planets as assumed in our model). P-R drag causes small grains to migrate inwards producing a flat radial distribution of small grains interior to the planetesimal disc \citep[e.g.][]{Wyatt2005b, Kennedy2015prdrag, Rigley2020}. However, the bulk of the distribution of large mm-sized grains will remain co-spatial with the parent planetesimals and thus P-R drag would not affect the $r^2$ surface density scaling of grains traced at mm wavelengths. 

Finally, our model assumes a pre-stirred disc or at least that the collisional timescale of the largest planetesimals is much longer than the stirring timescale throughout the disc. This may not be the true in the planet- or self-stirring scenarios depending on the system parameters \citep{Mustill2009, Kenyon2008, Krivov2018selfstirring}. If the stirring timescale was much longer, then the surface density just interior to the critical radius would have a much steeper profile \citep{Kennedy2010}. \cite{Marino2017a} fitted a similar radial profile model (allowing for a long stirring timescale) to ALMA observations of $\eta$~Corvi assuming the disc is being self-stirred, finding a that in order to reproduce the sharp inner edge the collisional timescale had to be shorter than the stirring timescale. However, self-stirring was found to be unlikely to explain the best-fit values as they required very small planetesimals that would be unlikely to have stirred the disc. Future work could evaluate this type of self-stirred models in a systematic way to a larger sample of systems and use more up-to-date self-stirring timescales \citep[e.g.][]{Krivov2018selfstirring}, which could change the conclusions of \cite{Marino2017a}.

\section{Conclusions} 
\label{sec:concl}

In this paper we have presented an analytical model for the collisional evolution of debris discs considering a three-phase size distribution and we showed how it can be applied to interpret the morphology of debris discs at mm wavelengths. In contrast to previous and similar analytic models, here we particularly focused on how collisional evolution is expected to shape the inner edge of a disc forcing the surface density to increase with radius to the power of 2 out to a critical radius. We use this model to derive simple analytical equations to constrain the total surface density of solids and maximum planetesimal size based on quantities that can be derived from observations, such as the dust surface density and the disc critical radius where the slope of the surface density flattens. 

We tested if this simple collisional model is consistent with ALMA observations of seven wide debris discs: HD~107146, q$^1$eri, HR~8799, AU~Mic, 49~Ceti, HD~206893, and HD~92945. We do this in a two step process using both parametric and non-parametric models to constrain the location and sharpness of the disc inner edge. We first used \textsc{frank} to fit the visibility data of each disc and derive a non-parametric model to determine an approximate shape of the disc at a higher resolution than conventional imaging techniques. We then used an MCMC to fit a parametric model to the visibilities and estimate the inner surface density slope. Based on those values, we determined if they are consistent with collisional evolution or if truncation by planets was more likely.

For 4 out of the 7 discs (HR~8799, HD~206893, 49~Ceti, and AU~Mic) we found the inner edges are consistent with a power-law index of 2, i.e. consistent with the models for collisional evolution. For those we found that the inner edge location could be explained by low disc masses relative to a Minimum Mass Solar Nebula and small planetesimals. In fact, we found for HR~8799, 49~Ceti and AU~Mic that the largest planetesimals could be sub-km in size. This is because these discs are large and young, and therefore do not require large planetesimals to sustain their dust levels. While the presence of large planetesimals is not strictly required to explain the dust levels in these systems, their presence cannot be excluded.

For the remaining three discs we found that the inner edges were sharper than predicted by collisional evolution, and thus they must have been set by something else. We explored the possibility that the inner edges were set by the interaction with planets (even for those with shallow edges), and we derived minimum planet masses to carve the inner edges within the age of the systems. For single planets carving the inner edge we found values of between 0.2 and 2 $\rm M_{\rm Jup}$. For multi-planet systems carving the inner edge we found masses between 1 and 70 $\rm M_{\oplus}$. We also derived the minimum planet mass for stirring the disc through secular interactions, and we found masses ranging between 0.6 $\rm M_{\oplus}$ and 2000 $\rm M_{\oplus}$. All of these values are much lower or at least consistent with detection limit of direct imaging observations from ground-based instruments, except for HR~8799. However, JWST could detect some of these; all of these systems will be observed during cycle 1. Such observations will be able to test some of our predictions and provide further insights about how the inner edge of these discs was shaped. 

Finally, during the non-parametric modelling of HD~107146, we discovered that there was an extra gap in the disc. This was present in archival data, but impossible to see in clean images. New higher resolution images and our parametric modelling confirmed this finding. The double gap morphology was recovered from all data sets with \textsc{frank} and was also what the parametric model fits converged towards. This highlights the importance of non-parametric fits like \textsc{frank} to extract detailed radial information prior to fitting a parametric model. Cycle 1 JWST observations of this disc will search for low mass planets interior and in between the disc and thus  provide strong constraints on the origin of these gaps.

\section*{Acknowledgments}

Throughout this project, Sebastian Marino was supported by a Junior Research Fellowship from Jesus College, University of Cambridge, and currently by a Royal Society University Research Fellowship. Amaia Imaz Blanco is thankful for the University of Cambridge Institute of Astronomy summer program, as without it I would never have gotten involved in this project that I have so enjoyed. G.M.K. is supported by the Royal Society as a Royal Society University Research Fellow. This paper makes use of the following ALMA data: ADS/JAO.ALMA\#2012.1.00198.S, ADS/JAO.ALMA\#2015.1.01260.S, ADS/JAO.ALMA\#2015.1.00307.S, ADS/JAO.ALMA\#2016.1.00104.S, ADS/JAO.ALMA\#2016.1.00195.S, ADS/JAO.ALMA\#2016.1.00907.S, ADS/JAO.ALMA\#2017.1.00167.S, ADS/JAO.ALMA\#2017.1.00467.S, ADS/JAO.ALMA\#2017.1.00825.S, ADS/JAO.ALMA\#2017.1.00828.S and ADS/JAO.ALMA\#2019.1.00189.S. ALMA is a partnership of ESO (representing its member states), NSF (USA), and NINS (Japan), together with NRC (Canada), MOST and ASIAA (Taiwan), and KASI (Republic of Korea), in cooperation with the Republic of Chile. The Joint ALMA Observatory is operated by ESO,
AUI/NRAO, and NAOJ. VF acknowledges funding from the National Aeronautics and Space Administration through the Exoplanet Research Program under Grant No. 80NSSC21K0394 (PI: S. Ertel). S.P. acknowledges support from FONDECYT 1191934 and ANID -- Millennium Science Initiative Program -- Center Code NCN2021\_080.


\section*{Data Availability}

The data underlying this article will be shared on reasonable request to the corresponding author. The ALMA data are publicly available and can be queried and downloaded directly from the ALMA archive at \href{https://almascience.nrao.edu/asax/}{https://almascience.nrao.edu/asax/}. {\sc Frankenstein} is publicly available at https://github.com/discsim/frank.



\bibliographystyle{mnras}
\bibliography{lib} 



\appendix

\section{New HD~107146 data}
\label{app:data}

New ALMA data was acquired as part of cycle 7 for the project 2019.1.00189.S (PI: S. Marino). The new observations were in band 7 (0.87~mm) and aimed to image the continuum at a much higher resolution ($\sim0.2\arcsec$) than previous observations to resolve the known gap in this system \citep{Marino2018b}. The cycle 7 observations included observations at a more compact configuration to be able to recover the large structure as this disc is $\sim10\arcsec$ in diameter. The compact configuration observations were obtained in December 2019 in two execution blocks, and this data subset was used in the analysis performed by \cite{Marino2021}. Due to the shutdown of ALMA in 2020 during the COVID-19 pandemic, only $\sim20\%$ of the extended configuration observations were carried out in May 2021. Nevertheless, these new observations proved useful in our analysis and thus we included them.

The spectral setup for both sets of observations was set in time division mode, with four spectral windows with a low spectral resolution to image the continuum emission. Their central frequencies were 336.5, 338.4, 348.5, and 350.5 GHz, with 128 channels each and a bandwidth of 2~GHz. Calibration of the raw data was done using the ALMA pipeline with \textsc{CASA} version 5.6.1-8 for the compact configuration data and with 6.2.1.7 for the extended configuration data, which included the flagging of two antennas. In addition, we flagged antenna DV08 after consultation with the helpdesk to reduce imaging artefacts. After flagging, there was a total of 41 antennas available with minimum and maximum baselines of 15 and 312~m in the compact configuration, and 39 antennas with minimum and maximum baselines of 15 and 1400~m in the extended configuration. When imaged separately, the compact and extended configuration clean images have an rms of 17 and 26 $\mu$Jy~beam$^{-1}$ with Briggs weights (robust=0.5).

\begin{figure}
\centering
\includegraphics[width=1.0\columnwidth]{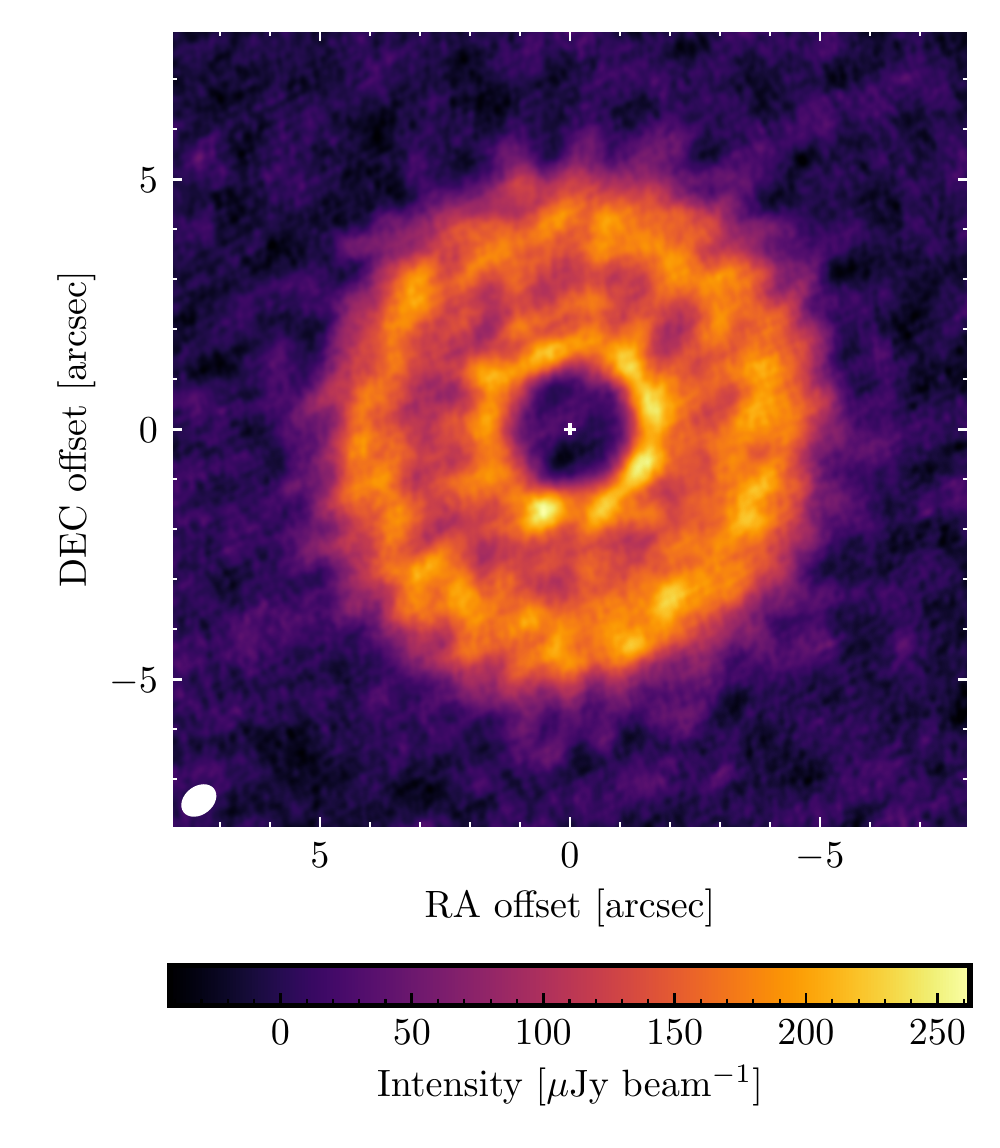}
\includegraphics[width=1.0\columnwidth]{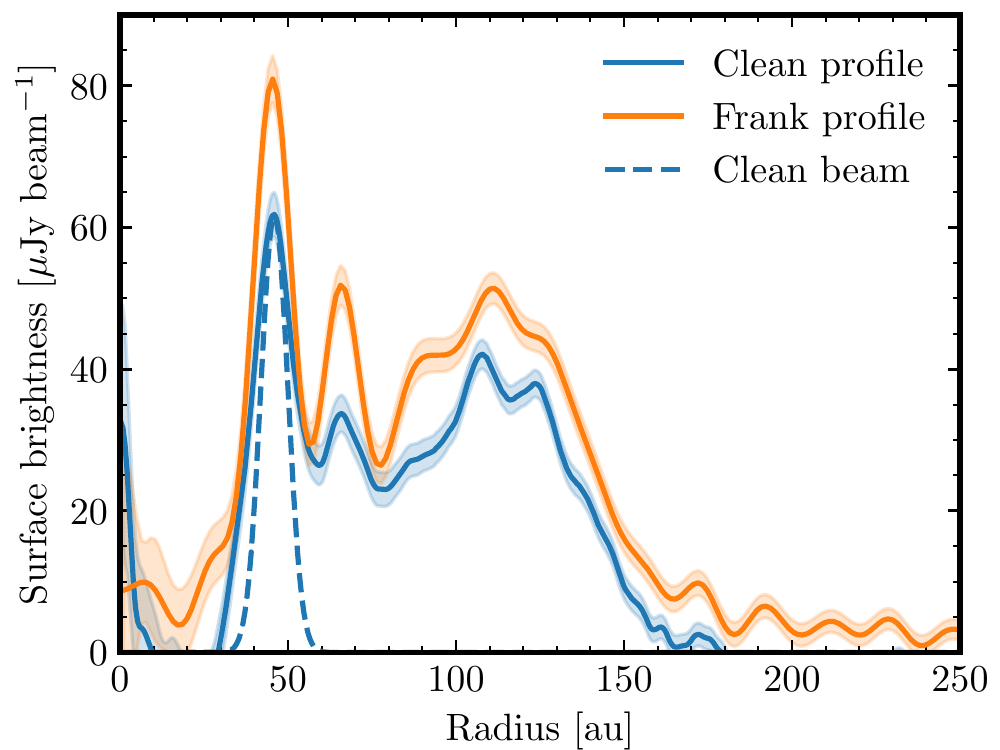}
    \caption{{\it Top panel:} Clean continuum image of HD~107146 from the new cycle 7 band 7 data. This image corresponds to Briggs weights (robust=1.0), giving a rms of 14$\mu$Jy~beam$^{-1}$ and a beam size of $0\farcs77\times0\farcs57$. {\it Bottom panel:} Deprojected intensity profiles obtained from a clean image with Briggs weights and robust=0.3 (blue) vs the intensity profile reconstructed from \textsc{frank} (orange). The shaded regions represent $\pm1\sigma$ confidence levels.}
    \label{fig:hd107146_data}
\end{figure}

The previous analysis done of the compact configuration data by \cite{Marino2021} revealed that the inner emission discovered in \cite{Marino2018b} was a background submillimeter galaxy. Due to HD~107146's proper motion, the relative position of this galaxy has changed over time. Therefore, prior to combining the data of both configurations, we subtract this source using a 2D Gaussian according to the best fit from \cite{Marino2021} at its corresponding position in December 2019 and May 2021. In addition, we use the CASA task fixplanets to change the coordinates of the phase center of the compact configuration observations (without changing the $uv$ coordinates or visibilities. Finally, we combine the two data sets and image the visibilities with tclean. The resulting image with Briggs weights (robust=1.0) is presented in the top panel of Figure~\ref{fig:hd107146_data}. This image reconstruction is dominated by the compact configuration data. The bottom panel shows the deprojected and azimuthally averaged emission of the disc (obtained from a clean image with a robust parameter of 0.3) together with the intensity profile derived by \textsc{frank} using all available band 7 data. The new data confirms the finding the double gap structure found by \textsc{frank} in the old band 6 and 7 data, although the two dips are not recovered with the same amplitude due to the poorer resolution of the Clean image ($\sim0.3\arcsec=8$~au). Moreover, the clean image seems to be missing some flux likely due to the low weights given to the short baselines through a robust parameter value of 0.3. This flux is well recovered with our parametric model in \S\ref{sec:fits}.

\section{Parametric Models}

\label{app:parametervalues}

Table \ref{tab:allparameters} presents the best fit values of all the parameters described in \S\ref{sec:fits} that we fit to the binned visibilities.

\begin{table*}
\caption{Best fit parameters for each of the models based on the MCMC results. The best fit value and uncertainties are based on the 16th, 50th and 84th percentiles of the marginalised probability distribution. If a parameter distribution reached our prior boundaries we instead report a 5\% lower or a 95\% upper limit (i.e. $2\sigma$.} Where a certain parameter was not used for a particular system, this cell is left blank and if a parameter was unconstrained this is labelled by a U. 
\label{tab:allparameters}
\begin{tabular}{@{}cccccccc@{}}
\toprule
Parameter & HR~8799 & q$^1$eri & HD~92945 & HD~107146 & HD~206893 & 49~Ceti & AU~Mic \\ \midrule
${\rm f_{\rm disc6}} [\rm mJy]$ & ... & $6.1^{+0.3}_{-0.3}$ & ... & $15.35^{+0.14}_{-0.13}$ & $0.88^{+0.05}_{0.05}$ & ...  & $4.88^{+0.07}_{-0.07}$ \\
${\rm f_{\rm disc7}} [\rm mJy]$ & $7.2^{+0.6}_{-0.5}$ & $13.2^{+0.5}_{-0.4}$ & $9.8^{+0.4}_{-0.4}$ & $29.1^{+0.3}_{-0.3}$ & $2.5^{+0.2}_{-0.2}$ & ... & ... \\
${\rm f_{\rm disc8} [mJy]} $ & ... & ...  & ... & ... & ... & $36^{+3}_{-3}$ & ... \\
${\rm f_{\star6}} [\rm mJy]$ & ... & $0.058^{+0.014}_{-0.014}$ & ... & $0.022^{+0.006}_{-0.006}$ & $0.013^{+0.005}_{-0.006}$ & ... & $0.23^{+0.02}_{-0.02}$ \\
${\rm f_{\star7} [mJy]} $ & $0.070^{+0.012}_{-0.012}$ & $0.161^{+0.015}_{-0.015}$ & $0.04^{+0.02}_{-0.02}$ & $0.04^{+0.01}_{-0.01}$  & $0.041^{+0.010}_{-0.010}$ & ... &...  \\
${\rm f_{\star8} [mJy]} $ & ... & ...  & ... & ... & ... & $<0.42$ & ...  \\
$r_{\rm t}$ [au] & $237^{+11}_{-11}$ & $84^{+1}_{-1}$ & $54^{+2}_{-2}$ & $44^{+2}_{-2}$ & $34.8^{+6.5}_{-9.6}$ & $131^{+13}_{-12}$ & $36.4^{+0.7}_{-0.7}$ \\
$\eta$ & $<1.6$ & $>2.8$ & U & $2.8^{+1.2}_{-0.7}$ & U & $<6.7$ & U \\
$\alpha_{\rm i}$ & $1.7^{+0.3}_{-0.3}$ & $4.2^{+0.5}_{-0.4}$ & $7^{+2}_{-2}$ & $6.7^{+0.9}_{-0.7}$ & $>0.55$ & $0.8^{+0.4}_{-0.3}$ & $0.9^{+0.4}_{-0.4}$ \\
$\alpha_{\rm m}$ & ... & ... & $-1.3^{+0.4}_{-0.6}$ & $-0.7^{+0.1}_{-0.2}$ & $0.4^{+0.6}_{-0.8}$ & ... & ... \\
$\alpha_{\rm o}$ & $-4.4^{+0.4}_{-0.5}$ & $-3.14^{+0.09}_{-0.10}$ &...  & ... & ... & $-3.5^{+0.4}_{-0.5}$ & $-9.9^{+1.0}_{1.4}$ \\
$r_{\rm out}$ [au] & ... & ... & $133^{+5}_{-7}$ & $144.3^{+0.9}_{-1.1}$ & $120^{+20}_{-20}$ & ... & ... \\
$l_{\rm out}$ [au] & ... & ... & $23^{+7}_{-5}$ & $19^{+1}_{-1}$ & $44^{+7}_{-6}$ & ... & ... \\
$\delta_1$ & ... & ... & $0.66^{+0.11}_{0.09}$ & $0.69^{+0.10}_{-0.09}$ & $0.92^{+0.05}_{-0.08}$ & ... & $0.94^{+0.04}_{-0.08}$ \\
$r_1$ [au] & ... & ... & $72.0^{+1.5}_{-1.5}$ & $56.0^{+0.7}_{-0.6}$ & $69^{+3}_{-3}$ & ... & $17.1^{+1.2}_{-1.4}$ \\
$\sigma_1$ [au] & ... & ... & $8^{+4}_{-4}$ & $3.3^{+0.7}_{-0.5}$ & $17^{+4}_{-4}$ & ... & $5^{+2}_{-1}$ \\
$\delta_2$ & ... & ... & ... & $0.60^{+0.05}_{-0.03}$ & ... & ... & ... \\
$r_2$ [au] & ... & ... & ... & $78.3^{+1.1}_{-1.2}$ & ... & ... & ... \\
$\sigma_2$ [au] & ... & ... & ... & $18^{+3}_{-2}$ & ... & ... & ... \\ \bottomrule
\end{tabular}
\end{table*}

\section{Inner Edge Slope Derivation for Intensity Profiles}
\label{app::inneredgederivation}
The derivation begins with equation \ref{eq:intensity} where $B(T)$ is the blackbody equation, $T$ is the temperature and $\tau$ is the optical depth. 

\begin{equation}
\label{eq:intensity}
    I=B(T)\times(1-\exp{(-\tau)})
\end{equation}
In the Rayleigh-Jean regime, which debris discs are, the black-body equation correlates with temperature as in Equation \ref{eq:blackbody}
\begin{equation}
    B(T,\lambda\gg\lambda_c)\propto T
    \label{eq:blackbody}
\end{equation}

The optical depth, $\tau$, is equal to the surface density of solids ($\Sigma$) times the opacity ($\kappa$), which together with the assumption that $\tau\ll1$ (which is the case for debris discs), leads to 
\begin{equation}
    I\propto T(r)\tau=T(r)\kappa\Sigma(r)
    \label{eq:intensityproportionality}
\end{equation}
Assuming the dust temperature is equal to the equilibrium temperature we have $ T(r)\propto r^{-1/2}$. Within the disc critical radius we expect $ \Sigma\propto r^\gamma$. Therefore, we find 
\begin{equation}
    \implies I\propto r^{-\frac{1}{2}} \kappa \Sigma (r) \implies I\propto r^{-1/2+\gamma}
    \label{eq:Iprop}
\end{equation}
This means that $\alpha_{\rm i}$ (the intensity power law index) is $\gamma-1/2$.


\bsp	
\label{lastpage}
\end{document}